\documentclass[twocolumn,aip,jcp,reprint,superscriptaddress,longbibliography]{revtex4-2}%

\usepackage{comment}
\usepackage{graphicx}
\usepackage{physics}
\usepackage{color}
\usepackage{tabularx}
\usepackage{multirow}
\usepackage{dcolumn}
\usepackage{longtable}
\usepackage{tensor}
\usepackage[utf8]{inputenc}
\usepackage[english]{babel}
\usepackage[T1]{fontenc}
\usepackage[hidelinks]{hyperref}
\usepackage{bm}
\usepackage{amsmath}
\usepackage{amsfonts}
\usepackage{amssymb}
\usepackage{adjustbox}
\usepackage{textcomp, gensymb}
\providecommand{\U}[1]{\protect\rule{.1in}{.1in}}

\newcommand{\fig}[1]{Fig.~\ref{#1}} %
\def\be{\begin{equation}} %
\def\ee{\end{equation}} %
\newcommand{\bea}{\begin{eqnarray}}
\newcommand{\eea}{\end{eqnarray}}

\newcommand{\BC}{\textcolor{black}}

\newcommand{\hatU}{\hat U}
\newcommand{\hatH}{\hat H}

\begin{document}

\title{Multistate iterative qubit coupled cluster (MS-iQCC): a quantum-inspired, state-averaged approach to ground- and excited-state energies}

\author{Robert A. Lang}
\affiliation{Chemical Physics Theory Group, Department of Chemistry,
  University of Toronto, Toronto, Ontario, M5S 3H6, Canada}
\affiliation{Department of Physical and Environmental Sciences,
  University of Toronto Scarborough, Toronto, Ontario, M1C 1A4,
  Canada}

\author{Shashank G. Mehendale}
\affiliation{Chemical Physics Theory Group, Department of Chemistry,
  University of Toronto, Toronto, Ontario, M5S 3H6, Canada}
\affiliation{Department of Physical and Environmental Sciences,
  University of Toronto Scarborough, Toronto, Ontario, M1C 1A4,
  Canada}
  
\author{Ilya G. Ryabinkin}
\affiliation{OTI Lumionics Inc., 3415 American Drive Unit 1, Mississauga, Ontario L4V 1T4, Canada}

\author{Artur F. Izmaylov}
\email{artur.izmaylov@utoronto.ca}
\affiliation{Chemical Physics Theory Group, Department of Chemistry,
  University of Toronto, Toronto, Ontario, M5S 3H6, Canada}
\affiliation{Department of Physical and Environmental Sciences,
  University of Toronto Scarborough, Toronto, Ontario, M1C 1A4,
  Canada}

\date{\today}

\begin{abstract}
We introduce the multistate iterative qubit coupled cluster (MS-iQCC) method, a quantum-inspired algorithm that runs efficiently on classical hardware and is designed to predict both ground and excited electronic states of molecules. Accurate excited-state energetics are essential for interpreting spectroscopy and chemical reactivity, but standard electronic structure methods are either too computationally expensive for larger systems or lose reliability in the presence of strong electron correlation. MS-iQCC addresses this challenge by simultaneously optimizing multiple electronic states in a single, state-averaged procedure that treats ground and excited states on equal footing. This removes the energetic bias that is introduced when excited states are computed one at a time and constrained to remain orthogonal to previously optimized states. The approach supports multireference electronic structure by allowing multideterminantal initial guesses and by adaptively building a compact exponential ansatz from a pool of qubit excitation generators. We apply MS-iQCC to H$_4$, H$_2$O, N$_2$, and C$_2$, including strongly correlated geometries, and observe robust convergence of all targeted state energies to chemically meaningful accuracy across their potential energy surfaces.

\end{abstract}

\maketitle

\section{Introduction}

The accurate determination of ground and excited electronic states is essential for understanding and predicting molecular properties and reactivity. Many physical and chemical phenomena depend critically on excited-state energetics and potential energy surfaces\cite{Domcke2011, Yarkony2012}. Classical computational chemistry provides a variety of approaches for these tasks, but their computational cost grows rapidly with system size. In particular, while multiconfigurational and perturbative methods such as MC-SCF\cite{Roos1980, Olsen1988, Malmqvist1990, LiManni2011}, CASPT2\cite{Andersson1990, Andersson1992, RoosAndersson1996}, and DMRG\cite{ChanSharma2011, Chandross1999, HuChan2015, Freitag2020, Barford2001} can provide highly accurate results, they become prohibitively expensive for larger molecules. More computationally affordable methods, such as time-dependent density functional theory (TDDFT) \cite{sarkar2021_tddft_benchmark,froitzheim2024_tddft_tadf,Genin2022}, offer broader applicability but often lack the accuracy required to reliably describe complex excited-state phenomena. Therefore, in this work, we focus on developing an alternative approach motivated by quantum computational concepts but executable efficiently on classical hardware.

Recent developments in quantum algorithms, especially the variational quantum eigensolver (VQE) \cite{Peruzzo2014, McClean_2016vqe, Cerezo2021, TILLY20221} and qubit coupled cluster (QCC) \cite{iQCC,iQCC_ILC} frameworks, have demonstrated new ways to approximate molecular eigenstates. However, near-term quantum devices (NISQ) remain limited by coherence times and circuit depth constraints, making it challenging to implement deep circuits required for accurate state preparation\cite{StatePrep} and measurement\cite{QMeas}. Consequently, quantum-inspired approaches—algorithms leveraging quantum formalisms but implemented on classical computers—offer an attractive intermediate step.
Yet, another alternative that became popular recently is quantum sampling-based diagonalization (QSD) techniques\cite{Barison_2025,qsd2025}. 
QSD uses an unoptimized unitary ansatz to produce Slater determinants 
by sampling the quantum state on a quantum computer. These Slater determinants are used to build a subspace for solving the Hamiltonian eigenvalue problem on a classical computer.   

Our method, the multistate iterative qubit coupled cluster (MS-iQCC), belongs to this class of quantum-inspired techniques. Although originally motivated by the variational quantum eigensolver and iterative qubit coupled cluster (iQCC) \cite{iQCC} algorithms, MS-iQCC is designed to run entirely on classical hardware. \BC{Recently, the iQCC method has been deployed using strictly classical compute to solve electronic structure problems up to 100 electrons in 100 molecular orbitals, corresponding to 200 qubit problems \cite{200iQCC}.} The multistate variant presented in this work retains the adaptive, exponential ansatz structure of iQCC but extends it to simultaneously describe multiple electronic states in a state-averaged, unbiased manner. This makes MS-iQCC particularly suited for classical computation today, while remaining directly relevant to future fault-tolerant quantum computing.

The adaptive exponential ansatz scheme used in MS-iQCC ensures separability for non-interacting fragments—an essential property shared with coupled cluster theory \cite{Helgaker}—but combines this with the variational flexibility of modern adaptive approaches\cite{ADAPT,qAdapt}. This separability also makes MS-iQCC conceptually preferable to selected configuration interaction methods \cite{cipsi_original,heatbath_ci,adaptive_ci,shci_modern}. Quantum excited-state methods such as the variational quantum deflation (VQD) \cite{Higgott2019, WenYv2021} and orthogonal state reduction variational eigensolver (OSRVE) \cite{Xie2022} suffer from accumulation of errors that arises from sequential optimization of excited states.

\BC{State-specific excited-state solvers, such as $\Delta$ADAPT-VQE\cite{delta_ADAPT_VQE}, address this challenge by extending the $\Delta$SCF method\cite{delta_SCF} to the quantum-computing setting. By employing a state-specific Hamiltonian and a non-Aufbau reference state, these approaches can target excited states without first computing the ground state. While this is advantageous, the algorithm must be executed separately for each target state. In addition, to prevent variational collapse to the ground state, one typically employs the maximum overlap method, which requires an initial guess sufficiently close to the desired target orbitals\cite{Delta_SCF_MOM}.}
        
\BC{In contrast, state-averaged methods optimize the average energy of a set of orthogonal states, which is always bounded from below by the average of the lowest eigenvalues. Consequently, a single execution of the algorithm can yield multiple excited states. Moreover, state-averaged approaches treat all target eigenstates on an equal footing, unlike state-specific methods. This strategy has been adopted in recent extensions of ADAPT-VQE to excited states, including ADAPT-VQE-SCF \cite{ADAPT_VQE_SCF}, which combines state-averaged ADAPT-VQE with orbital optimization, and the subsequent MORE-ADAPT-VQE\cite{Grimsley_2025}, which is similar in spirit but omits orbital optimization. As with many VQE-based proposals, applying these methods to industrially relevant problem instances may require a large number of circuit repetitions to evaluate energies and gradients. Moreover, by adaptively evolving the Hamiltonian rather than the reference states, MS-iQCC has the additional advantage of using an unrestricted operator pool in constructing the variational ansatz, whereas the aforementioned approaches rely on a predefined operator pool.}

Finally, while MS-iQCC is designed for efficient classical computation, it naturally produces compact unitary transformations suitable for preparing approximate eigenstates. These transformations can serve as high-quality inputs for future fault-tolerant quantum phase estimation (QPE) \cite{Aspuru2005,Griffiths1996, Lin_Tong_2022, Wang_Franca_2023, Ding_Lin_2023,Gratsea2024} algorithms, enabling accurate and scalable excited-state calculations once quantum hardware matures.

The rest of this paper is organized as follows. Section \ref{sec:theory} details the theoretical formulation of MS-iQCC. Section \ref{sec:numerics} presents numerical demonstrations for several molecular systems, including H$_4$, H$_2$O, N$_2$, and C$_2$. Finally, Section \ref{sec:conclusions} summarizes the main findings and discusses directions for future development.

\section{Theory} \label{sec:theory}

\subsubsection{The MS-iQCC formalism}
Given a Hamiltonian $\hat{H}^{(0)}$ written in terms of qubit operators, and a set of orthogonal reference states $\{\ket{I}\}_{I =1}^{N_s}$, the goal of MS-iQCC is to iteratively construct a parametrized unitary operator $\hat{U}$ such that, 
\begin{align}
    \hat{H}_{qcc} &= \hat{U}\hatH^{(0)}\hatU^\dagger, \label{eq: H_qcc}\\
    \hat{H}_{qcc}\ket{I} &= E_I\ket{I}, 
\end{align}
 where $\hat{H}_{qcc}$ is the resulting Hamiltonian from MS-iQCC algorithm, and $E_I$ is the $I^\text{th}$ lowest eigenenergy of $\hatH^{(0)}$. Equivalently, in the Schrödinger picture, $\hatU$  maps the reference state $\ket{I}$ to the $I^\text{th}$ lowest eigenstate of $\hatH^{(0)}$. The reference states $\ket{I}$ could potentially be multi-reference in nature, which is necessary when targeting excited multiplet states, \cite{Helgaker}  or excited singlets with open shell character. \cite{BALKOVA1992}\\

\begin{figure} 
    \includegraphics[width=0.48\textwidth]{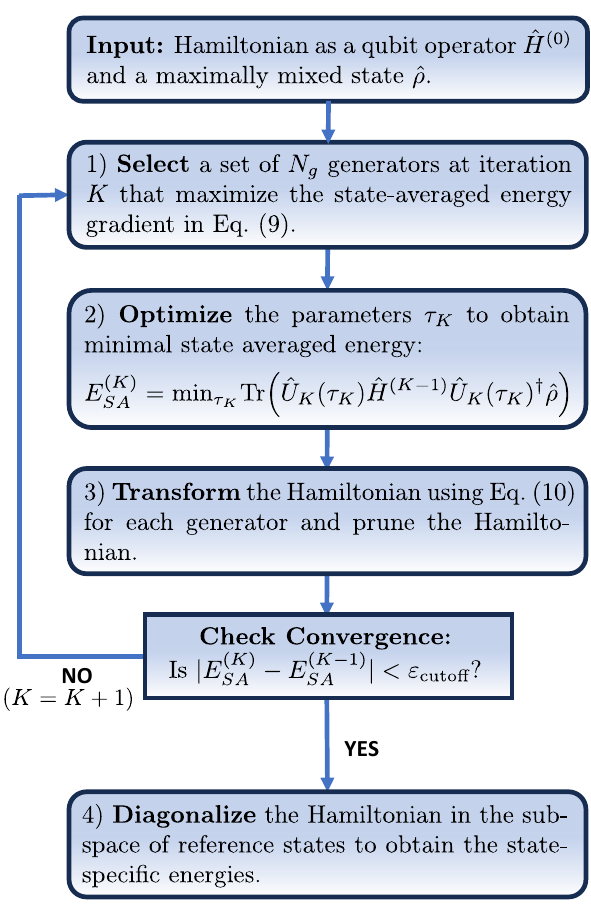}
    \vspace{-2em}
    \caption{ A step-by-step description of the MS-iQCC algorithm. The input mixed state $\hat{\rho}$ is defined in Eq. \eqref{eq: input mixed state}. The diagonalization in step 4 marks the end of the algorithm and the resulting Hamiltonian is the $\hat{H}_{qcc}$ defined in Eq. \eqref{eq: H_qcc}. The diagonalization can also be performed at the end of each iteration to track the convergence of state-specific energies instead of state-averaged energy. If one chooses $N_g = 1$, then step 4 becomes irrelevant and the whole procedure reduces to GS-iQCC. \vspace{-1em}}
    \label{fig:flowchart}
\end{figure}

 $\hatH^{(0)}$ is obtained by applying a fermion-to-qubit mapping to the second-quantized molecular Hamiltonian. The MS-iQCC procedure then starts with constructing a density operator out of reference states $\{\ket{I}\}_{I =1}^{N_s}$,
 \begin{align}
     \hat{\rho} = \sum_{I=1}^{N_s} w_I \ketbra{I}, \label{eq: input mixed state}
 \end{align}
where $w_I$ are probabilities. To avoid state-specific bias we set all $w_I = 1/N_s$. The underlying principle behind MS-iQCC is the state-ensemble variational principle.\cite{gvp} 
At any iteration $K$, a parametrized unitary transformation is applied to the Hamiltonian from the previous iteration, $\hatH^{(K-1)}$, to yield
\begin{align}
    \hat{\mathcal{H}}^{(K)}(\tau_K) = \hatU_K(\tau_K)\hatH^{(K-1)}\hatU_K(\tau_K)^{\dagger}, \label{eq: general transformed Hamiltonian}
\end{align}
which gives the state-averaged energy
\begin{align}
    E_{SA}(\tau_K) = \text{Tr}\Big(\hat{\mathcal{H}}^{(K)}(\tau_K)\hat{\rho}\Big), \label{eq: general st avg energy}
\end{align}
where $\tau_K$ is a parameter vector. $E_{SA}(\tau_K)$ is then minimized as a function of the parameters to yield the optimal parameter vector $\tau^*_K$ and the optimal Hamiltonian at iteration $K$, $\hatH^{(K)} = \hat{\mathcal{H}}^{(K)}(\tau^*_K)$. Because of the variational principle, we can be certain that
\begin{align}
    E_{SA}(\tau^*_K) \geq \dfrac{1}{N_s}\sum_{I=1}^{N_s} E_I \label{eq:gvp}
\end{align}
which is the average of the true $N_s$ lowest energies of the Hamiltonian $\hat{H}^{(0)}$. Thus by adding a suitably chosen unitary $\hatU_K$ at each iteration, we can bring $E_{SA}(\tau^*_K)$ closer to its lower bound. A subtlety unique to MS-iQCC (absent in GS-iQCC) is that, even when equality in Eq. \eqref{eq:gvp} holds, the state-averaged value $\expval{\hatH^{(K)}}{I}$ need not necessarily equal $E_I$. This is a direct consequence of choosing all $w_I = 1/N_s$, since any unitary that acts non trivially only on the reference subspace, leaves the density operator $\hat{\rho}$, and hence the state-averaged energy $E_{SA}$, unchanged. In order to get the state-specific energy at the end of an iteration, the Hamiltonian $\hatH^{(K)}$ needs to be diagonalized within the subspace spanned by the reference states $\{\ket{I}\}_{I =1}^{N_s}$. The algorithm can be terminated when $\Delta E_{SA} = |E_{SA}^{(K)} - E_{SA}^{(K-1)}|$ falls below a threshold. The overview of the algorithm is given in \fig{fig:flowchart}. \\

One of the main advantages of MS-iQCC is in the choice of unitaries $\hatU(\tau_K)$. At each iteration $K$, one chooses a set of $N_g$ unitaries generated by the basis elements of the $\mathfrak{su}(2^{N_q})$ algebra. These basis elements are the $4^{N_q} -1$ possible $N_q$-qubit ``Pauli terms'' omitting the $N_q$-qubit identity. Pauli terms are defined as the $N_q$-fold tensor products of the Pauli operations $\hat x$, $\hat y$, $\hat z$, along with the single qubit identity,
\begin{align}
\hat T_{\alpha} = \bigotimes_{p=1}^{N_q} \hat \sigma_{p}^{(\alpha)}, \quad \hat \sigma \in \{\hat x, \hat y, \hat z, \hat 1 \}
\end{align}
where $\hat \sigma_p$ is a single qubit operation $\hat \sigma$ applied to qubit $p$. A layer of $N_g$ generators is expressed as
\begin{align}
\hat U_K (\boldsymbol{\tau}_{K}) = \prod_{\alpha=1}^{N_g} \exp(-i \tau_{\alpha}^{(K)} \hat T_{\alpha}^{(K)} / 2),
\end{align}
where $\boldsymbol{\tau}_{K} = (\tau_1^{(K)}, \tau_2^{(K)}, \hdots, \tau_{N_g}^{(K)})$ are $N_g$ real variational parameters referred to as \textit{amplitudes.} \\

Due to the exponential number of possible generators to include in $\hat U_K$, the MS-iQCC algorithm, much like the GS-iQCC, relies on heuristic approaches to find $N_g$ generators which will have the most impact on variationally lowering the energy. We use the magnitude of the energy derivative at $\tau_{\alpha} = 0$ as an importance measure; the procedure selects $N_g$ generators with the largest $g_\alpha$ given by,
\begin{align}
g_\alpha & =  \left| \frac{\partial}{\partial \tau_{\alpha}} \Tr(e^{i \tau_{\alpha} \hat T_{\alpha} / 2 } \hat H^{(K-1)} e^{- i \tau_{\alpha} \hat T_{\alpha} / 2 } \hat \rho  ) \Big|_{\tau_{\alpha}=0}    \right|, \nonumber \\
& = \dfrac{1}{N_s}\left| \sum_{I=1}^{N_s} \mathrm{Im} \left( \bra{I} \hat H^{(K-1)} \hat T_{\alpha} \ket{I} \right)   \right|. \label{eq:gradient_maintext}
\end{align}
Critically, the algorithm does \textit{not} employ a predefined operator pool from which candidate Pauli terms are screened. This fact differentiates the algorithm from other adaptive schemes, which employ operator pools restricted to polynomially scaling sizes to facilitate efficient generator screening. Instead, the effective Hamiltonian $\hat H^{(K-1)}$ directly informs us on the structure of Pauli terms $\hat T_{\alpha}^{(K)}$ with non-zero gradient. From Eq. (\ref{eq:gradient_maintext}), a screening algorithm efficiently checks which $\hat T_{\alpha}^{(K)}$ generate excited configuration $\ket{I_\alpha} \equiv \hat T_{\alpha}^{(K)} \ket{I}$ which can be coupled to $\ket{I}$ through $\hat H^{(K-1)}$. This is reminiscent of the selection criterion used in heat-bath CI (HCI).\cite{Holmes2016, Sharma2017} The space generated by the screened Pauli operators $\hat T_\alpha$ is called the ``Direct Interaction Space'' (DIS). Due to its importance, we dedicate Section \ref{sec:ms_dis} to describe its construction, and provide further details in Appendix \ref{app:dis_mr}.\\

Another important feature of the algorithm presented here is the ability to classically obtain $\hatH^{(K)}$ from $\hatH^{(K-1)}$ efficiently, which relies on the self-inverse property of any Pauli term;
\begin{align} \label{eq:pauli_dressing}
e^{i \tau_\alpha \hat T_{\alpha} / 2} \hat H e^{- i \tau_\alpha \hat T_{\alpha} / 2} = & \hat H - \frac{i}{2} \sin(\tau_\alpha) [\hat H, \hat T_\alpha] \nonumber \\&  + \frac{1}{2}\left( 1 - \cos(\tau_\alpha) \right) \left( \hat T_\alpha \hat H \hat T_\alpha - \hat H  \right).
\end{align}

By repeating this for each of the $N_g$ generators and over $K$ iterations, the number of Pauli terms in the resulting Hamiltonian $\hatH^{(K)}$ formally scales as $O((3/2)^{N_gK})$.\cite{iQCC} In practice, the proliferation of terms in the Hamiltonian is often drastically below  the theoretical scaling.\cite{Genin2022} Numerically, the proliferation of terms is mitigated by `compressing' the effective Hamiltonian after each application of Eq. (\ref{eq:pauli_dressing}). As in iQCC, Pauli terms with coefficient magnitude less than a compression threshold $\varepsilon_c$ are pruned. \BC{This idea has also appeared in recent works related to sparse Pauli dynamics \cite{SPD}. Optionally, Pauli terms could be pruned based on other criteria, such as fermionic rank of the corresponding Majorana products \cite{majorana_products}}.
\vspace{-1.6em}

\subsubsection{Multi-state direct interaction space} \label{sec:ms_dis}

To understand the construction of DIS for MS-iQCC, we start with the simpler case of single determinant reference states, $\{\ket{I}\}{_{I=1}^{N_s}} = \{\ket{\phi_I}\}_{I=1}^{N_s}$. The case of multi-configurational reference states is considered in Appendix \ref{app:dis_mr}. The ensemble is written as
\begin{align*}
    \hat{\rho} = \dfrac{1}{N_s}\sum_I \ketbra{\phi_I}.
\end{align*}
Before maximizing the gradient in Eq. \eqref{eq:gradient_maintext}, we can eliminate many Pauli words from consideration as they will certainly yield zero gradient. Start by writing the Hamiltonian in Ising factorized form \cite{iQCC},
\begin{align}
    \hat{H} = \sum_i \eta_i \hat{P}_i = \sum_j \hat{D}_j \hat{X}_j, \label{eq:H_ising}
\end{align}
where $\hat P_i$ is a Pauli term, and $\hat{D}_j = \sum_k \eta^{(j)}_k\hat{Z}^{(j)}_k$, with $\eta_k^{(j)}$ as real numbers, $\hat{X}$ and $\hat{Z}$ as tensor products of strictly $\hat{x}$ and $\hat{z}$ operations respectively (up to inclusion of identity). For real valued electronic Hamiltonians such as those in the absence of external magnetic field, the Hamiltonian can be equivalently written as $\hat{H} = \sum_j \hat{X}_j\hat{D}_j$. We also express Pauli word $\hat T_\alpha$ as $\hat T_\alpha = \theta_\alpha \hat{X}_\alpha\hat{Z}_\alpha$, where $\theta_\alpha \in \{1, -1, i, -i\}$. An important point to note here is that $\theta_\alpha$ is imaginary if and only if there is an odd number of Pauli $\hat{y}$ operators in the tensor product defining $\hat T_\alpha$. This implies that the operators $\hat{x}$ and $\hat{z}$ defining $\hat{X}_\alpha$ and $\hat{Z}_\alpha$ overlap on odd number of qubits. In such case, we say that $\hat{X}_\alpha$ and $\hat{Z}_\alpha$ have \textit{odd overlapping support}.\\

With this representation of the Hamiltonian and Pauli words, the gradient can be rewritten as,
\begin{align*}
    g_\alpha & = \dfrac{1}{N_s}\left| \sum_{I=1}^{N_s} \mathrm{Im} \left( \bra{\phi_I} \sum_j\hat{D}_j\hat{X}_j \cdot \theta_\alpha \hat{X}_\alpha \hat{Z}_\alpha \ket{\phi_I} \right)   \right|.
\end{align*}
As $\ket{\phi_I}$ are single Slater determinants, we can reduce this expression to
\begin{align}
    g_\alpha & = \dfrac{|\mathrm{Im}(\theta_\alpha)|}{N_s}\ \left| \sum_{I=1}^{N_s} \lambda^{(\alpha)}_I \Xi_I^{(\alpha)} \right|, \label{eq:g_sa}
\end{align}
where  $\Xi_I^{(\alpha)} = \sum_jD_j^{(I)}\expval{\hat{X}_j\hat{X}_\alpha}{\phi_I}$  with\\ $D_j^{(I)} = \expval{\hat{D}_j}{\phi_I}$, and $\lambda^{(\alpha)}_I = \expval{\hat{Z}_\alpha}{\phi_I}$. Based on Eq. \eqref{eq:g_sa}, we can identify $\hat T_\alpha$'s that do not contribute to the gradient as follows.
\begin{itemize}
    \item $g_\alpha = 0$ if $\hat{X}_\alpha$ and $\hat{Z}_\alpha$ have even overlapping support, as this would imply $\mathrm{Im}(\theta_\alpha) = 0$.
    \item $g_\alpha = 0$ if $\hat{X}_\alpha \neq \hat{X}_j$ for any $j$, as this would imply $\Xi_I^{(\alpha)} = 0$ for all $I$.
\end{itemize}

This is the key result underlying the use of an unrestricted pool of unitary generators in MS-iQCC. Procedurally, the DIS is constructed by first choosing $\hat{X}_\alpha = \hat{X}_j$ for each $j$ from Eq. \eqref{eq:H_ising}, followed by finding $\hat{Z}_\alpha$ such that $\hat{X}_\alpha$ and $\hat{Z}_\alpha$ have odd overlapping support. For these Pauli words, the gradient can be simplified to,
\begin{align}
    g_{\alpha} & = \dfrac{1}{N_s}\left| \sum_{I=1}^{N_s} \lambda^{(\alpha)}_I D_\alpha^{(I)} \right|. \label{eq:z_align}
\end{align}
where $\hat{D}_\alpha$ equals one of the $\hat{D}_j$ in Eq. \eqref{eq:H_ising}, and $\lambda^{(\alpha)}_{I}$ depends on the choice of $\hat{Z}_\alpha$. We call maximization of Eq. \eqref{eq:z_align} the \textit{phase-alignment problem}: for each fixed $\hat{D}_\alpha$ choose $\hat Z_{\alpha}$ to align the phases $\lambda_I ^{(\alpha)} = \pm 1$ so as to maximize $g_\alpha$. The problem amounts to an $N_q$-dimensional binary optimization problem and solving for the optimal solution is equivalent to finding the global optimum to a weighted MAX-SAT problem.\cite{Mills2000} Appendix \ref{app:phase_align} details the phase-alignment problem, presenting an optimal approach (\texttt{OPT}) and a greedy approximation (\texttt{GreedySAT}). Note, unlike MS-iQCC, GS-iQCC would have a single term instead of a sum in Eq. \eqref{eq:z_align}, which would imply that $\lambda_I^{(\alpha)}$ is an irrelevant global phase, avoiding the problem of phase-alignment entirely. In Appendix \ref{app:dis_mr}, we extend the DIS construction to state-averages of generally multiconfigurational reference states. We show that this general DIS construction is conceptually similar to the one presented in this section, with gradients retaining the general form of Eq. (\ref{eq:g_sa}), but with modification of the $\Xi^{(\alpha)}$ components.

\section{Numerical results} \label{sec:numerics}

In this section we perform simultaneous ground and excited state calculations using the MS-iQCC procedure on a set of modestly sized, albeit strongly correlated, chemical systems. Prior to their fermion-to-qubit encodings, all second quantized Hamiltonians were obtained in the restricted Hartree Fock (RHF) orbital basis. All energy errors reported throughout are calculated via the absolute difference of the state-specific energy and the target state FCI/CASCI energy computed via the \texttt{PySCF} library.\cite{pyscf} The Hamiltonian growth factors reported throughout refer to the proportionality in number of Pauli terms between an iQCC effective Hamiltonian and the initial qubit-mapped Hamiltonian. The $\hat H$ growth factor at iteration $K$ is 
\begin{align}
G_K = \frac{M_K}{M_1},
\end{align}
where $M_K$ and $M_1$ are the number of Pauli terms in the $K^{\rm{th}}$ iteration effective and initial Hamiltonians, respectively. We also report the error in target state fidelities (overlaps) defined as:
\begin{align}
F_{I} = 1 - \left | \bra{E_I} \hat U_{\rm{iQCC}} \ket{I} \right |^2, \label{eq: overlap_error_defn}
\end{align}
computed using the \texttt{Openfermion} library  \cite{openfermion}, to showcase convergence of states along with energies, which could be of independent interest to the community. We use a compression threshold to numerically suppress the proliferation of terms in a transformed iQCC Hamiltonian. Utilizing a threshold of $\varepsilon_c$ involves truncating the effective Hamiltonian as 
\begin{align}
\hat H^{(K)}_{\varepsilon_c} = \sum_{|\eta_{i}| \geq \varepsilon_c} \eta_i^{(K)} \hat P_i^{(K)},
\end{align}
where such a truncation is performed directly following each step of single-generator dressing.

When describing RHF electronic configurations, we represent them as Fock occupation vectors in the spin orbital basis. The fermionic modes are enumerated by ascending orbital energy, e.g., the HF configuration for an arbitrary system is expressed as $\ket{1, \hdots, 1, 0, \hdots, 0}$.\\

\subsection{H$_4$} \label{sec:h4}

\begin{figure*}[!htb] 
    \centering
    \includegraphics[width=0.9\textwidth]{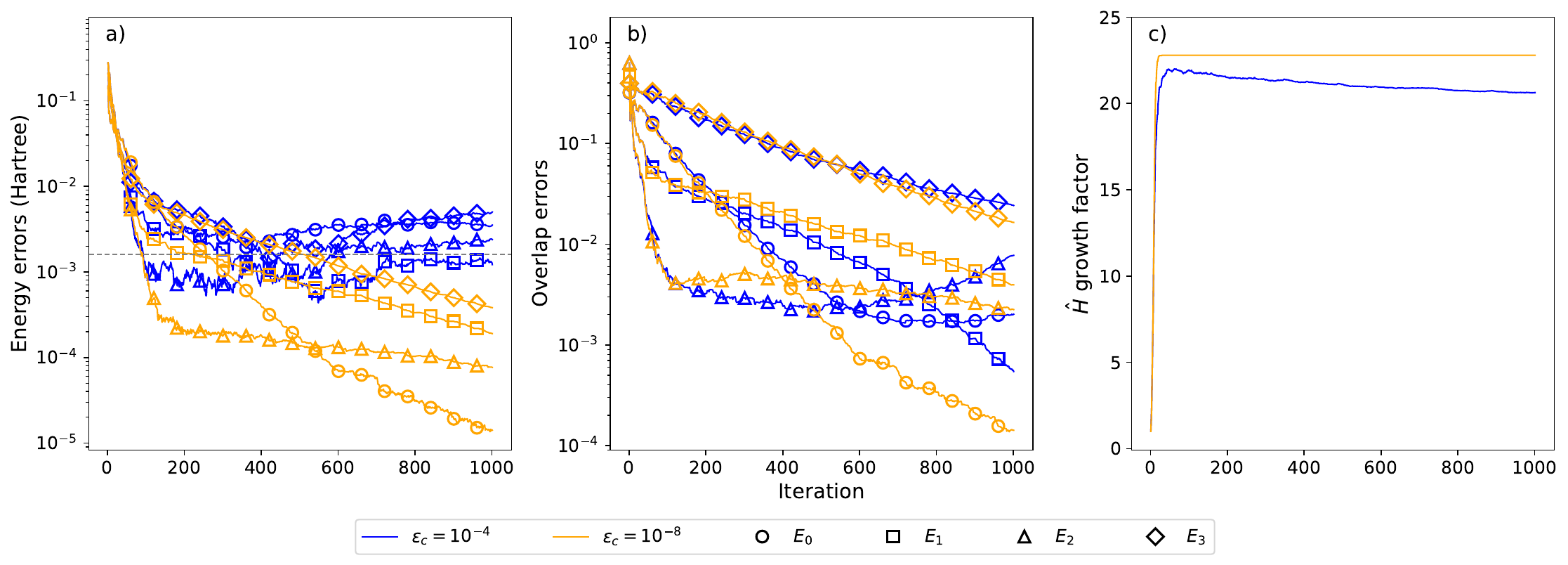}
    \caption{MS-iQCC applied to the determination of the four lowest energy eigenstates for the H$_4$ chain at an equidistant separation of $2r_e$ in the STO-3G basis set. The $\varepsilon_c$ denotes the compression threshold used. a) Errors in the state-specific energies relative to the associated FCI target energies. b) Errors in the squared wavefunction overlaps of the iQCC-rotated trial states with their corresponding target eigenstates (see Eq. \eqref{eq: overlap_error_defn}). c) The ratio of number of terms in the effective iQCC Hamiltonian and the initial qubit-mapped Hamiltonian.}
    \label{fig:h4_ext}
\end{figure*}

Herein, MS-iQCC is applied to multiple state determination for the linear equidistant H$_4$ chain in the STO-3G basis set. We assess the algorithm in determining the four lowest lying states of the molecule with equidistant H$-$H separation of twice the equilibrium, $2r_e = 1.9$ \AA{}. The qubit Hamiltonian is obtained through the Jordan-Wigner map, resulting in an $N_q = 8$ qubit Hamiltonian. The MS-iQCC algorithm is employed in determining the four lowest lying eigenstates. A model space of $L=8$ configurations was employed, consisting of the Fock vectors
\begin{align}
\ket{\phi_1} & = \ket{1 1 1 1 0 0 0 0}, \ket{\phi_2} = \ket{1 1 0 0 1 1 0 0}, \ket{\phi_3} = \ket{1 1 1 0 0 1 0 0}, \nonumber \\  
\ket{\phi_4} & = \ket{1 1 0 1 1 0 0 0},   \ket{\phi_5} = \ket{1 0 1 1 0 1 0 0}, \ket{\phi_6} = \ket{0 1 1 1 1 0 0 0}, \nonumber \\  
\ket{\phi_7} & = \ket{0 0 1 1 1 1 0 0}, \ket{\phi_8} = \ket{1 1 0 0 0 0 1 1}. \label{eq:h4_model_space}
\end{align}
The multiconfigurational references are obtained by diagonalizing the $8$-dimensional Hamiltonian matrix in the subspace of Eq. (\ref{eq:h4_model_space}) and picking the four lowest eigenstates. Energetic errors, target state fidelity errors, and Hamiltonian growth factors for the  $N_s = 4$ state determination are presented in \fig{fig:h4_ext}. We perform the multi-state determination using $N_g = 1$ generator per iteration, and include the results for two different compression thresholds, $\varepsilon_c$. In \fig{fig:h4_ext}, it is seen that an aggressive compression of $\varepsilon_c = 10^{-4}$ causes the state-specific energies to drift away at later iterations instead of converging to their target values. However, the target state fidelity trajectories over iterations are almost identical to those obtained using $\varepsilon_c = 10^{-8}$, with both compression thresholds providing all four state fidelities of $> 0.9$ by $K \approx 500$. For $\varepsilon_c = 10^{-8}$, all four state-specific energies achieve chemical accuracy after $\sim 500$ iterations of a single generator at a time. The effective iQCC Hamiltonian grows rapidly until iteration $K \approx 15$, at which point the Hamiltonians reach the so-called ``saturation point",\cite{iQCC, iQCC_ILC} where the set of Pauli terms in the Hamiltonian becomes algebraically closed with respect to commutator with the selected generators. Notably, the effective Hamiltonians using $\varepsilon_c = 10^{-4}$ exhibit a maximal number of terms in vicinity to when $\varepsilon_c = 10^{-8}$ achieves saturation, however the number of terms begins to slightly decrease past this point. The rapid early saturation, and near-saturation of the aggressively compressed Hamiltonians, can be attributed to the low Hilbert space dimensionality. The scaling of effective Hamiltonians in the MS-iQCC procedure along with the effect of compression is revisited for the larger problems of N$_2$ and C$_2$ in Sections \ref{sec:n2} and \ref{sec:cr2} respectively.

\subsection{H$_2$O} \label{sec:h2o}

\begin{figure*}[!htb] 
    \centering
    \includegraphics[width=1.0\textwidth]{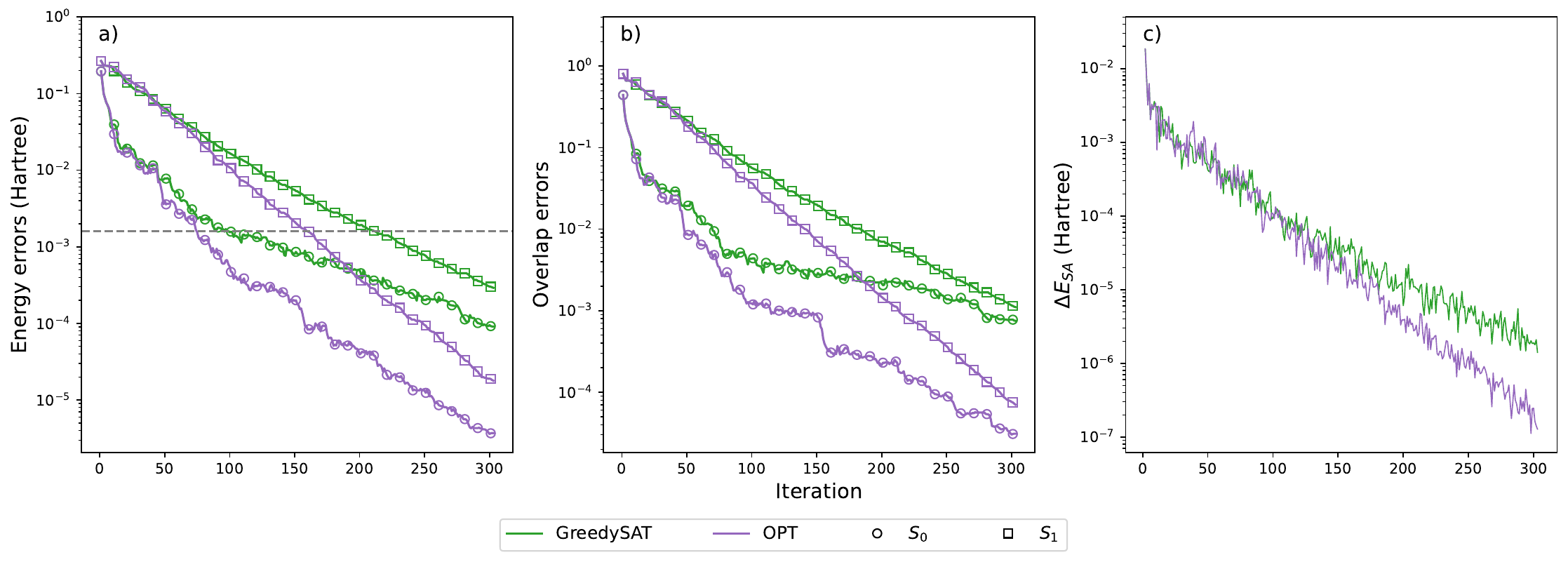}
    \caption{The MS-iQCC procedure with $N_g = 1$ applied to the simultaneous determination of the $\text{S}_0$ and $\text{S}_1$ states of the CAS($4$e, $4$o) model of stretched H$_2$O in the $6$-$31$G basis set. The $L=4$ configurations in Eq. (\ref{eq:h2o_model_space}) were used to define the model space. The global optimization described in Appendix \ref{sec:selection_opt} was utilized for phase-alignment. a) The errors of the MS-iQCC $\text{S}_0$ and $\text{S}_1$ estimates taken with respect to the exact $\text{S}_0$ and $\text{S}_1$ solutions within the CAS($4$e, $4$o) space. b) Errors in the squared overlaps of the MS-iQCC trial states with respect to the associated target state. c) $\Delta E_{SA}$ at iteration $K$ is given by the difference between the optimized $K^{\rm{th}}$ and $(K-1)^{\rm{th}}$ MS-iQCC state-averaged energy.}
    \label{fig:h2o_smallbasis}
\end{figure*}
\begin{figure*}[!htb] 
    \centering
    \includegraphics[width=1.0\textwidth]{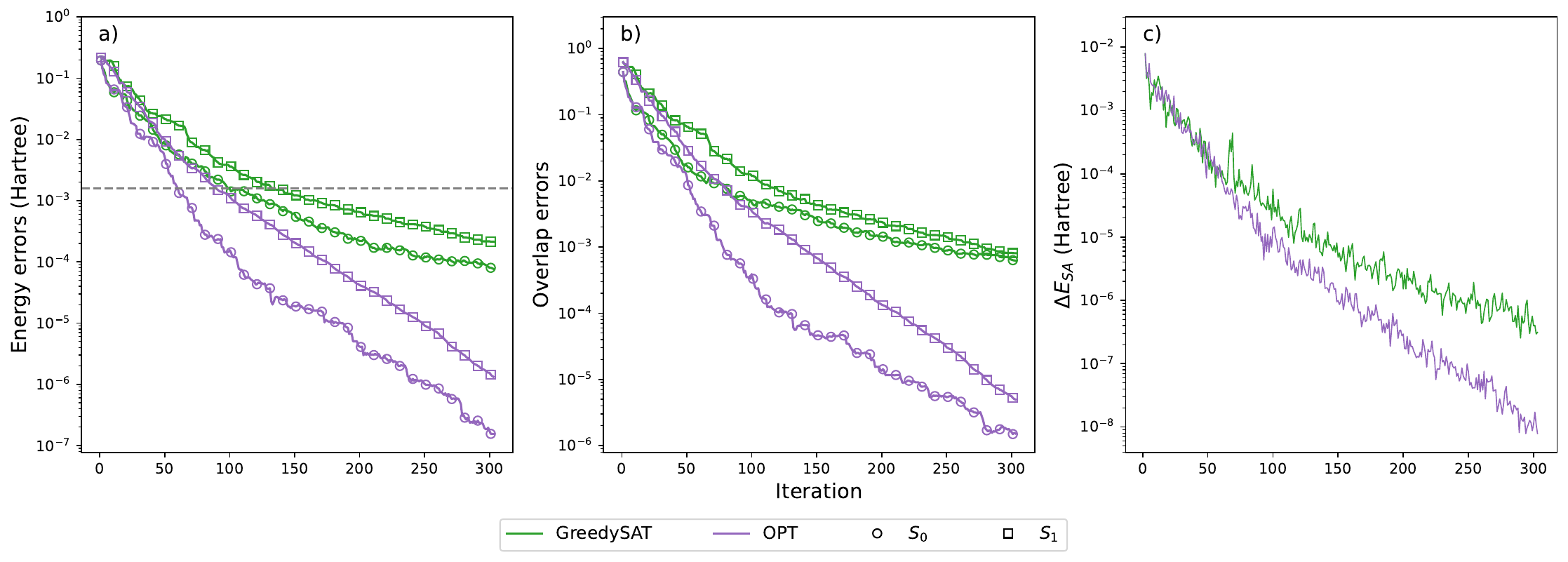}
    \caption{The same as \fig{fig:h2o_smallbasis}, however employing the slightly larger model space of dimension $K=6$, with extra configurations contributing to the $S_{1}$ trial 
 state reference given in Eq. (\ref{eq:h2o_model_space_ext}).}
    \label{fig:h2o_ext_basis}
\end{figure*}

Herein the MS-iQCC procedure is applied to a CAS($4$e, $4$o) model of the stretched H$_2$O molecule in the $6$-$31$G basis set. A distance of $2.35$ \AA{} was used for both O-H bonds, with a H-O-H bond angle of $107.6^\circ$. At such a geometry, the two lowest lying singlet states, $\text{S}_0$ and $\text{S}_1$, exhibit a high degree of multiconfigurational character, making their simultaneous estimation a challenging problem. In order to ensure the $\text{S}_1$ estimate does not converge to a lower energy high-spin solution, such as $\text{T}_1$, a spin-penalized Hamiltonian
\begin{align} \label{eq:H_spin_pen}
\hat H_s = \hat H + \mu \hat S^2
\end{align}
is utilized for generator screening and amplitude optimization, where $\mu = 0.25 \text{ a.u}$. We assess the performance of the MS-iQCC procedure utilizing only $N_g = 1$, i.e., a single generator per iteration, with varying reference states and phase-alignment procedures utilized. The MS-iQCC is applied to the $\text{S}_0$ and $\text{S}_1$ estimation using a model space comprised of a total of $L=4$ configurations,
\begin{align}
\ket{\phi_1} & = \ket{1 1 1 1 0 0 0 0}, \quad \ket{\phi_2} = \ket{1 1 0 0 1 1 0 0}, \nonumber \\ 
\ket{\phi_3} & = \ket{1 0 1 1 0 1 0 0}, \quad \ket{\phi_4} = \ket{0 1 1 1 1 0 0 0}. \label{eq:h2o_model_space}
\end{align}
The multiconfigurational references are obtained by diagonalizing the $4$-dimensional Hamiltonian matrix in the subspace of Eq. (\ref{eq:h2o_model_space}) and picking the two lowest eigenstates.
In \fig{fig:h2o_smallbasis}, we compare the two phase-alignment procedures (see Appendix \ref{app:phase_align}), the optimal \texttt{OPT} with the heuristic \texttt{GreedySAT} approach. We observe that the \texttt{OPT} phase-alignment yields accelerated convergence of the state-specific energies and state-averaged energy compared to the \texttt{GreedySAT} phase-alignment. Convergence to chemical accuracy for both singlet states was accomplished using \texttt{GreedySAT} with roughly $50$ additional iterations. The slower convergence can be attributed to the heuristic nature of \texttt{GreedySAT}, which may potentially miss the true maximal gradient generators.

Notably, both phase-alignment methods achieve $> 0.9$ state fidelities for $\text{S}_0$ and $\text{S}_1$ with $K \sim 100$, despite the initial $\text{S}_1$ overlap being merely $0.19$. To assess the benefit of increasing the model space, we perform an identical MS-iQCC calculation in \fig{fig:h2o_ext_basis}, except with an additional $2$ electronic configurations included in the $\text{S}_1$ reference state, given by
\begin{align}
\ket{\phi_5} = \ket{1 1 1 0 0 0 0 1}, \quad \ket{\phi_6} = \ket{1 1 0 1 0 0 1 0}.\label{eq:h2o_model_space_ext}
\end{align}
Diagonalization in the $6$-dimensional model space yields a multiconfigurational reference for the $\text{S}_1$ state with an improved initial target fidelity of $0.39$. It is seen that the MS-iQCC algorithm using either of the phase-alignment techniques requires $\sim 50$ fewer iterations to achieve state-specific energies within chemical accuracy, yielding $\sim 33 \%$ and $\sim 25 \%$ reductions for the \texttt{OPT} and \texttt{GreedySAT} phase-alignments respectively.

\subsection{N$_2$} \label{sec:n2}

\begin{figure*}[!htb]
    \centering
    \includegraphics[width=1.0\textwidth]{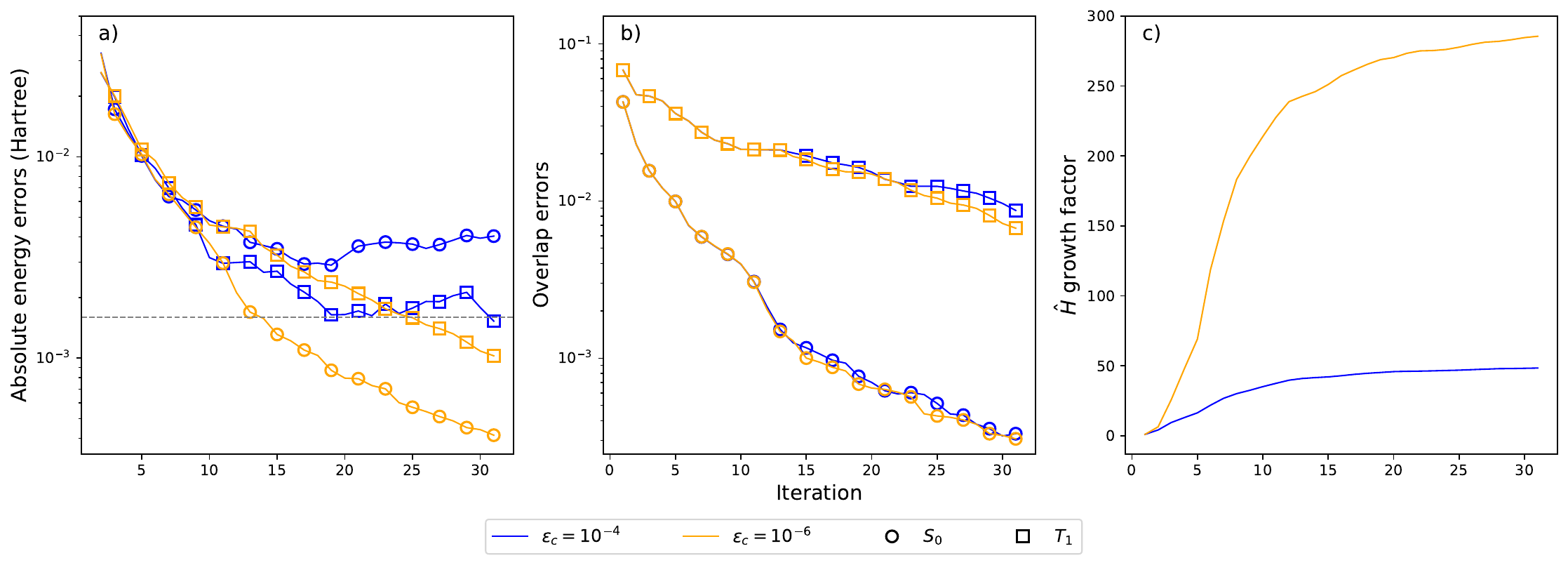}
    \caption{The MS-iQCC procedure applied to simultaneous determination of the $\text{S}_0$ and $\text{T}_1$ states of the CAS($6$e, $6$o) model of N$_2$ molecule at $r_e = 1.0975$ \AA{} bond distance in the STO-6G basis set. A total of $L=7$ computational basis states were used, with $N_g = 5$ generators used at each iteration, and \texttt{OPT} strategy utilized for phase-alignment. The subplots a, b, and c correspond to energy errors, fidelity errors, and the growth factors, respectively.}
     \label{fig:n2_eqbm}
\end{figure*}

\begin{figure*}[!htb]
    \centering
    \includegraphics[width=1.0\textwidth]{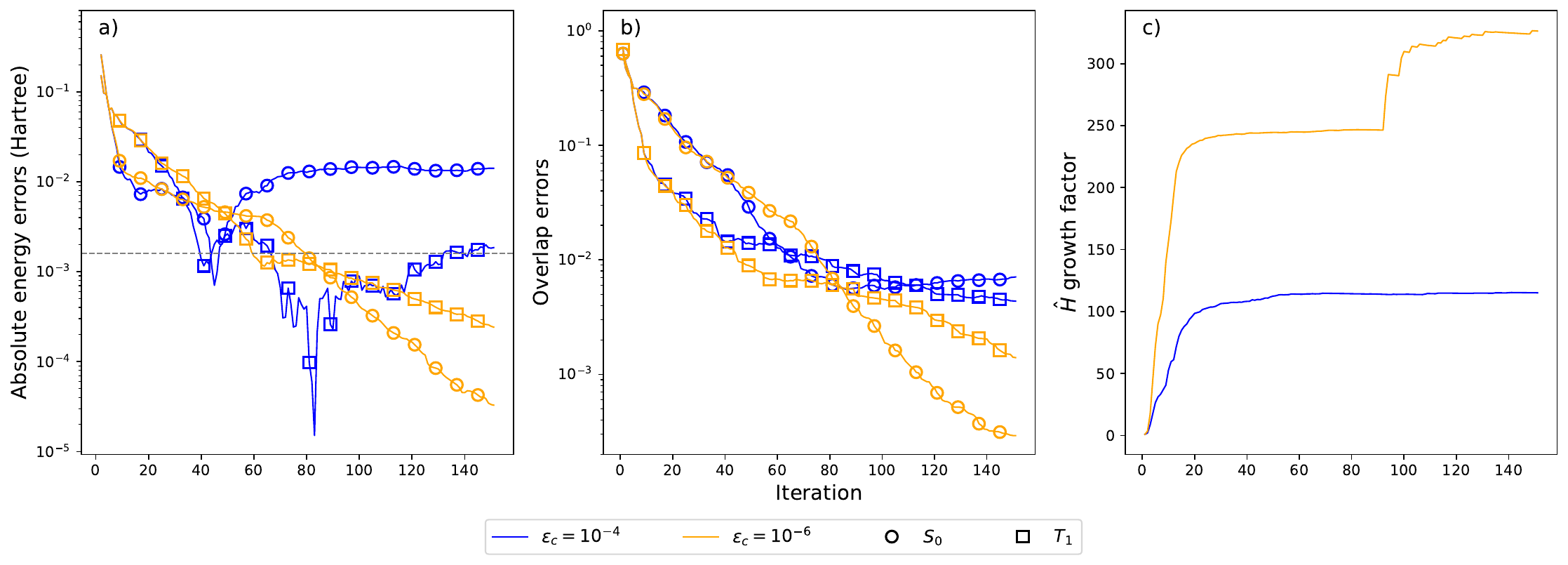}
    \caption{The same as \fig{fig:n2_eqbm} but performed at a N$-$N bond distance of $2r_e$.}
     \label{fig:n2_ext}
\end{figure*}

We assess the MS-iQCC procedure in performing simultaneous estimation of the ground ($\text{S}_0$) and first excited ($\text{T}_1$) states of N$_2$ in a CAS($6$e, $6$o) active space in the STO-6G basis set. We benchmark the algorithm at both equilibrium ($r_e=1.0975$ \AA{}) and extended ($2r_e$) bond distances. The qubit Hamiltonian is obtained under the parity mapping,\cite{parity} and a two qubit tapering procedure\cite{Bravyi_Temme:2017} is employed to obtain an $N_q = 10$ qubit Hamiltonian. To obtain the multiconfigurational reference states for $\text{S}_0$ and $\text{T}_1$, we utilize a small set of $L=7$ configurations from the singles-and-doubles manifold from Hartree-Fock. For the calculation performed at equilibrium geometry, we utilize electronic configurations with the following representation as Fock vectors
\begin{align}
\begin{split}
\ket{\phi_1} & = \ket{1 1 1 1 1 1 0 0 0 0 0 0},\ \ket{\phi_2} = \ket{0 0 1 1 1 1 1 1 0 0 0 0}, \\ 
\ket{\phi_3} & = \ket{1 1 0 0 1 1 0 0 1 1 0 0},\ \ket{\phi_4} = \ket{1 0 1 1 1 1 0 1 0 0 0 0}, \\ 
\ket{\phi_5} & = \ket{0 1 1 1 1 1 1 0 0 0 0 0},\ \ket{\phi_6} = \ket{1 1 1 0 1 1 0 0 0 1 0 0}, \\ 
\ket{\phi_7} & = \ket{1 1 0 1 1 1 0 0 1 0 0 0}.
\end{split} \label{eq:N2_eq_model_space}
\end{align}
At the extended bond length of $2r_e$, the model space used is 
\begin{align}
\begin{split}
\ket{\phi_1} & = \ket{1 1 1 1 1 1 0 0 0 0 0 0},\ \ket{\phi_2}  = \ket{1 1 0 0 1 1 1 1 0 0 0 0},  \\ 
\ket{\phi_3} & = \ket{1 1 1 1 0 0 0 0 1 1 0 0},\ \ket{\phi_4}  = \ket{1 1 1 0 1 1 0 1 0 0 0 0},  \\ 
\ket{\phi_5} & = \ket{1 1 0 1 1 1 1 0 0 0 0 0},\ \ket{\phi_6}  = \ket{1 1 1 1 1 0 0 0 0 1 0 0},  \\ 
\ket{\phi_7} & = \ket{1 1 1 1 0 1 0 0 1 0 0 0}.
\end{split} \label{eq:N2_2eq_model_space}
\end{align}
The choice of model space configurations was made by observing the highest weighted determinants in the $\text{S}_0$ and $\text{T}_1$ CISD solutions at both geometries, with $\ket{\phi_i}$ for $i \in \{ 1 ,2 ,3 \}$ being the top three contributing configurations to the $\text{S}_0$ solution, and the remaining four configurations coming from the $\text{T}_1$ solution. The Hamiltonian was diagonalized in the subspace of Eq. \eqref{eq:N2_eq_model_space} and Eq. \eqref{eq:N2_2eq_model_space} and the two lowest lying solutions were taken as the multiconfigurational reference states for $\text{S}_0$ and $\text{T}_1$ states. Figure \ref{fig:n2_eqbm} provides numerics for the MS-iQCC procedure applied to the two state determination at equilibrium geometry, using $N_g=5$ generators each iteration, for two different regimes of compression. Similarly to the situation seen for H$_4$, crudely compressing to a threshold of $10^{-4}$ is seen to provide unreliable convergence of state-specific energies, albeit providing target state fidelities on par with a compression of $10^{-6}$. The latter less severe compression is seen to converge both $\text{S}_0$ and $\text{T}_1$ trial energies to within chemical accuracy at $K=25$. Notably, the number of terms in the iQCC effective Hamiltonian is seen to be much smaller for $\varepsilon_c = 10^{-4}$ compared to $\varepsilon_c = 10^{-6}$. Such behaviour can be explained by the estimation at equilibrium geometry being dominated by dynamical/weak correlation, evident by the high overlaps of the starting references. In the non-strongly correlated regime, the optimized iQCC amplitudes $\tau_{\alpha}$ are generally small. From Eq. (\ref{eq:pauli_dressing}), new terms entering the updated Hamiltonian carry coefficient $\sin(\tau_{\alpha})$. As the Hamiltonian is iteratively dressed using small amplitudes, terms are being continually suppressed by $\sin(\tau_{\alpha}) \ll 1$ factors, leading to many terms being pruned during compression. \\

At the bond distance of $2r_e$, the reference states possess squared overlaps of $\sim 0.37$ and $\sim 0.31$  with the exact $\text{S}_0$ and $\text{T}_1$ states, respectively. The low overlaps of the reference states with their target states present this example as an interesting and challenging problem for simultaneous state estimation. Furthermore, following $\text{T}_1$, the next low-lying excited states are quintet and septet states. Since iQCC generators do not conserve spin symmetries, energetically quasidegenerate high spin states can be problematic. If the algorithm finds generators which maximally lower energy at early iterations, yet introduce large amounts of spin contamination, there is a risk of convergence to one such high spin eigenstate. A less severe scenario, yet still unfavorable, is that many iQCC iterations are required to approximately restore the desired spin quantum numbers to that of the lower-lying target state. Such behaviour is a clear manifestation of L\"{o}wdin's symmetry dilemma.\cite{Lykos1963} To avoid large amounts of spin contamination entering the trial states at early iterations, we again employ the spin-penalized Hamiltonian of Eq. (\ref{eq:H_spin_pen}) for operator screening and optimization, with a smaller penalty $\mu = 0.025$ a.u., to avoid overly penalizing triplet energies.

In \fig{fig:n2_ext}, numerics are provided for the $\text{S}_0$ and $\text{T}_1$ determinations at $2r_e$ bond length. Utilizing $\varepsilon_c = 10^{-6}$ is seen to provide systematic convergence towards the state-specific $\text{S}_0$ and $\text{T}_1$ energies, with chemical accuracy achieved by $K=80$ iterations. The poor fidelities of the reference states are rapidly improved, with both achieving $\sim 0.9$ squared overlap with their target states by $K=25$. Using $\varepsilon_c = 10^{-4}$ resulted in both the $\text{S}_0$ and $\text{T}_1$ energy estimates becoming non-variational, resulting in the sharp negative peaks in the trajectories of their absolute energy errors. The breaking of the variational behaviour is possible when compressing the effective Hamiltonian to low precision. Aggressively pruning terms of high coefficient magnitude can lead to significant spectral perturbations, and hence should be avoided when accurate energies are desired as the output of the MS-iQCC algorithm.

\subsection{C$_2$} \label{sec:cr2}

\begin{figure*}[!htb]
    \centering
    \includegraphics[width=\textwidth]{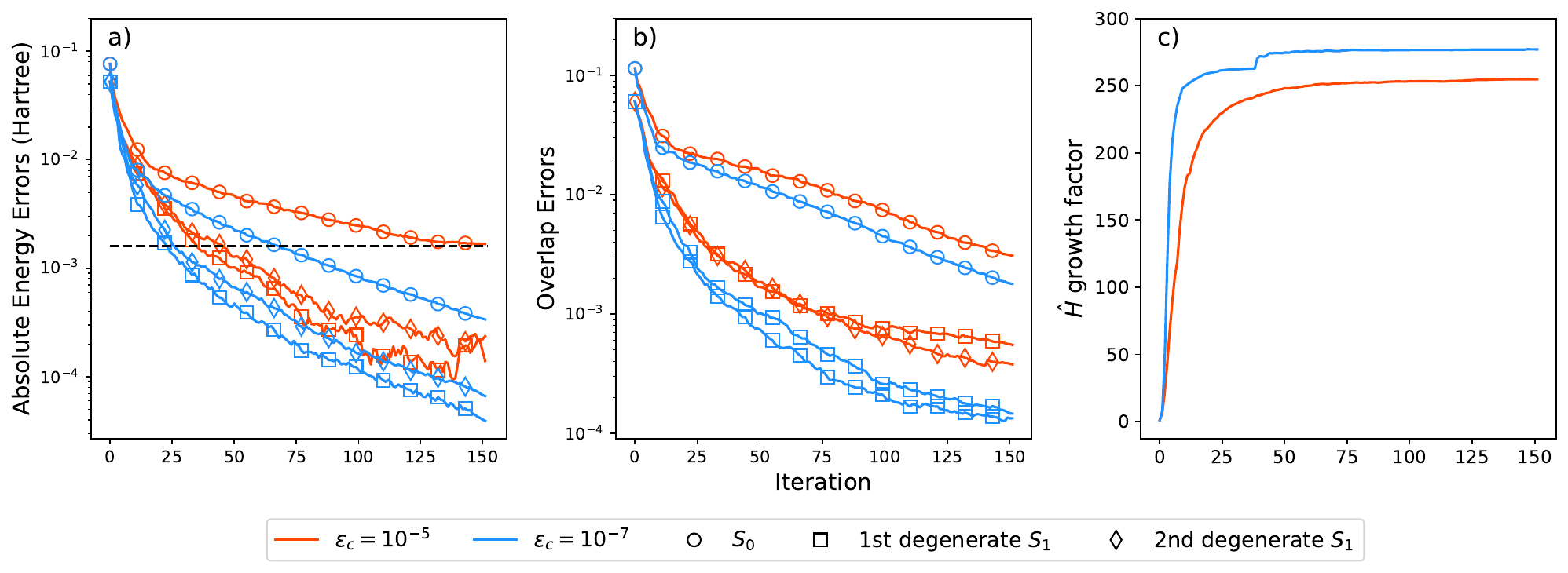}
    \caption{The MS-iQCC procedure applied to simultaneous determination of $\text{S}_0$ and the two degenerate $\text{S}_1$ states of the CAS($6$e, $6$o) model of C$_2$ molecule at $r_e = 1.2$ \AA{} bond distance in the cc-pVDZ basis. Each iteration utilized $N_g = 6$ generators, and \texttt{GreedySAT} strategy was utilized for phase-alignment. Two different compression factors were tested. The subplots a, b, and c correspond to energy errors, fidelity errors, and growth factors, respectively.}
     \label{fig:c2_re}
\end{figure*}

We demonstrate the usage of MS-iQCC procedure for simultaneous determination of S$_0$ state and two degenerate S$_1$ states of C$_2$ in the complete active space of 6 electrons and 6 orbitals of the cc-pVDZ basis, at a bond length of r$_e$ = 1.2 \AA{}. Here as well, the qubit Hamiltonian is obtained under the parity mapping, and a two qubit tapering procedure is employed to obtain an $N_q = 10$ qubit Hamiltonian. All target states have spin projection along the z-axis equal to zero. We choose the reference determinants by looking at the top contributors to the target states in the CISD space. They are given by,
\begin{align}
\ket{\phi_1} &= \ket{111111000000},\ \ket{\phi_2} = \ket{001111110000}, \nonumber \\
\ket{\phi_3} &= \ket{111101100000},\ \ket{\phi_4} = \ket{111110010000}, \nonumber \\
\ket{\phi_5} &= \ket{111011010000},\ \ket{\phi_6} = \ket{110111100000}. \nonumber
\end{align}

We diagonalize the spin penalized Hamiltonian of Eq. \eqref{eq:H_spin_pen} with $\mu = 0.025$ in the subspace spanned by these determinants and pick the three lowest lying states as the multiconfigurational references. We run the algorithm for 150 iterations with number of generators in each iteration set to 6. GreedySAT algorithm is utilized for solving the phase-alignment problem. \\

We study the results under two different compression thresholds. In \fig{fig:c2_re} a), we see that for $\varepsilon_c = 10^{-5}$, convergence of S$_0$ state to chemical accuracy is extremely slow, taking nearly 150 iterations. For S$_1$ states, chemical accuracy is reached within 50 iterations, although the convergence is substantially perturbed towards the end. This could be attributed to the choice of large compression threshold which might prune many important Pauli words off the Hamiltonian, breaking the variational nature of the problem. To recover a more systematic convergence for all states, we rerun the calculation with a smaller compression threshold of $\varepsilon_c = 10^{-7}$. We see that all three states converge systematically to the target energies and the chemical accuracy is reached around 25 iterations for excited states, and 75 iterations for the ground state. This suggests that the ground state is possibly more strongly correlated than the excited states, requiring more Slater determinants in its expression and hence more number of iQCC generators.\\

Unlike in the case of N$_2$ at equilibrium bond length (see \fig{fig:n2_eqbm}), we can see in \fig{fig:c2_re} b) that reducing the compression threshold improves the target state fidelities. In subplot c) Hamiltonian growth factor, as expected, displays a rise in number of terms as $\varepsilon_c$ is lowered. But compared to other molecules, we can see that the rise is less significant. The smaller compression threshold leads to a drastic rise in the growth factor at the beginning as most terms generated by unitaries will be retained, and the Hamiltonian terms become algebraically closed with respect to the dominant generators fairly quickly. In case of a more aggressive pruning, many iterations are needed to boost the coefficients of the important Pauli words to reach algebraic closure.

\section{Conclusions and outlook} \label{sec:conclusions}
In this work, we have introduced the MS-iQCC algorithm for simultaneous and unbiased determination of ground and excited state energies of qubit mapped Hamiltonians. We tested the algorithm on several strongly correlated molecules and found that, under moderate Hamiltonian compression thresholds, it demonstrated robust convergence of all state-specific energies to within chemical accuracy. In this regime, MS-iQCC functions as a fully classical multi-state solver. We have also shown that one can use spin penalized Hamiltonians to target excited states in different symmetry sectors. The efficacy of the MS-iQCC algorithm can be largely attributed to the unrestricted use of the full $\mathfrak{su}(2^{N_q})$ algebra acting on the $N_q$-qubit Hilbert space in selecting unitary generators. This results in an adaptively growing \textit{Hamiltonian}, in analogy to the use of an adaptively growing configuration space as used in iterative selected CI techniques.\cite{Holmes2016, Sharma2017}. Interestingly, for all the cases studied, the number of terms in the effective iQCC Hamiltonian converges rapidly, showing that the Pauli words in the Hamiltonian become algebraically closed with respect to the dominant generators being used. This property supports the observation that the growth in the number of terms of the iQCC Hamiltonian is substantially below the theoretical scaling \cite{iQCC}. We also showed that along with state specific energies, MS-iQCC displays a systematic convergence of target state fidelities. This could be of independent interest to the quantum computing community where state preparation is an important subroutine in extracting spectral properties using the Quantum Phase Estimation algorithm. Due to the compact nature of MS-iQCC unitaries and the ability to generate multiple excited states, it is worth exploring the benefits and limitations of using MS-iQCC as a quantum state preparation algorithm. \\

MS-iQCC has tunable parameters, such as the compression threshold and the number of generators per iteration, which control the trade-off between cost and accuracy. More consideration is needed in choosing their values, since their rigorous connection with target accuracy is not apparent. Another important component of the algorithm is the solution to phase-alignment problem. Since finding an optimal solution is prohibitively expensive for large systems, quality of greedy solutions strongly influences the number of iQCC iterations needed to reach a target accuracy. We believe further research is needed in improving these approximate solutions, or modifying steps in MS-iQCC that could suggest better strategies for tackling the phase-alignment problem.

\section{Acknowledgements}
We thank Junan Lin and Christian Schilling for stimulating discussions. A.F.I. acknowledges financial support from the Natural Sciences 
and Engineering Council of Canada (NSERC). This research was partly enabled by the support of Compute Ontario (computeontario.ca) and the Digital Research Alliance of Canada (alliancecan.ca). Part of the computations were performed on the Niagara and Trillium supercomputers at the SciNet HPC Consortium, and the NARVAL and RORQUAL supercomputers under the Calcul Quebec Consortium. SciNet is funded by Innovation, Science, and Economic Development Canada, the Digital Research Alliance of Canada, the Ontario Research Fund: Research Excellence, and the University of Toronto.

\bibliography{mrsaiqcc}

@STRING{Nature = {Nature}}

@STRING{Science = {Science}}

@article{Barison_2025,
   title={Quantum-centric computation of molecular excited states with extended sample-based quantum diagonalization},
   volume={10},
   ISSN={2058-9565},
   url={http://dx.doi.org/10.1088/2058-9565/adb781},
   DOI={10.1088/2058-9565/adb781},
   number={2},
   journal={Quantum Science and Technology},
   publisher={IOP Publishing},
   author={Barison, Stefano and Robledo Moreno, Javier and Motta, Mario},
   year={2025},
   month=feb, pages={025034} }

@misc{qsd2025,
      title={Quantum chemistry with provable convergence via randomized sample-based quantum diagonalization}, 
      author={Samuele Piccinelli and Alberto Baiardi and Max Rossmannek and Almudena Carrera Vazquez and Francesco Tacchino and Stefano Mensa and Edoardo Altamura and Ali Alavi and Mario Motta and Javier Robledo-Moreno and William Kirby and Kunal Sharma and Antonio Mezzacapo and Ivano Tavernelli},
      year={2025},
      eprint={2508.02578},
      archivePrefix={arXiv},
      primaryClass={quant-ph},
      url={https://arxiv.org/abs/2508.02578}, 
}

@article{sarkar2021_tddft_benchmark,
  title={Benchmarking TD-DFT and Wave Function Methods for Excitation Energies},
  author={Sarkar, R. and Casanova-Páez, M. and Neese, F. and Izsák, R.},
  journal={Journal of Chemical Theory and Computation},
  volume={17},
  number={11},
  pages={6612--6629},
  year={2021},
  doi={10.1021/acs.jctc.0c01228}
}

@article{froitzheim2024_tddft_tadf,
  title={Benchmarking Charge-Transfer Excited States in TADF Emitters: {$\Delta$}DFT Outperforms TD-DFT for Emission Energies},
  author={Froitzheim, T. and Kunze, L. and Grimme, S. and Herbert, J. M. and Mewes, J.-M.},
  journal={Journal of Physical Chemistry A},
  volume={128},
  number={28},
  pages={6324--6335},
  year={2024},
  doi={10.1021/acs.jpca.4c03654}
}

@article{cipsi_original,
  title={Configuration Interaction by Perturbatively Selected Iterative Method},
  author={Huron, Bernard and Malrieu, Jean-Pierre and Rancurel, Pierre},
  journal={The Journal of Chemical Physics},
  volume={58},
  number={12},
  pages={5745--5759},
  year={1973},
  doi={10.1063/1.1679199}
}

@article{heatbath_ci,
  title={An Efficient Selected Configuration Interaction Plus Perturbation Theory Algorithm},
  author={Holmes, Adam A. and Tubman, Norm M. and Umrigar, C. J.},
  journal={Journal of Chemical Theory and Computation},
  volume={12},
  number={8},
  pages={3674--3680},
  year={2016},
  doi={10.1021/acs.jctc.6b00407}
}

@article{adaptive_ci,
  title={Adaptive Configuration Interaction: A Multiconfigurational Wave Function Method Inspired by Machine Learning},
  author={Evangelista, Francesco A.},
  journal={The Journal of Chemical Physics},
  volume={140},
  number={12},
  pages={124114},
  year={2014},
  doi={10.1063/1.4869192}
}

@article{shci_modern,
  title={Semistochastic Heat-Bath Configuration Interaction Method: Selected Configuration Interaction with Perturbation Theory Using a Semistochastic Approach},
  author={Sharma, Sandeep and Holmes, Adam A. and Jeanmairet, Guillaume and Alavi, Ali and Umrigar, C. J.},
  journal={Journal of Chemical Theory and Computation},
  volume={13},
  number={4},
  pages={1595--1604},
  year={2017},
  doi={10.1021/acs.jctc.6b01081}
}

@article{StatePrep,
  title={A Quantum Computing View on Unitary Coupled Cluster Theory},
  author={Anand, A and Schleich, P and Alperin-Lea, S and Jensen, PWK and Sim, S and D{\'\i}az-Tinoco, M and Kottmann, JS and Degroote, M and Izmaylov, AF and Aspuru-Guzik, A},
  journal={Chem. Soc. Rev.},
  volume={51},
  pages={1659--1684},
  year={2022}
}

@article{QMeas,
  title={Quantum Measurement for Quantum Chemistry on a Quantum Computer},
  author={Patel, Smik and Jayakumar, Praveen and Yen, Tzu-Ching and Izmaylov, Artur F.},
  journal={Chemical Reviews},
  volume={125},
  pages={7490--7524},
  year={2025}
}

@article{parity,
  title={Fermionic quantum computation},
  author={Bravyi, Sergey B and Kitaev, Alexei Yu},
  journal={Annals of Physics},
  volume={298},
  number={1},
  pages={210--226},
  year={2002},
  publisher={Elsevier}
}

@article{Bravyi_Temme:2017,
  title={Tapering off qubits to simulate fermionic Hamiltonians},
  author={Bravyi, Sergey and Gambetta, Jay M and Mezzacapo, Antonio and Temme, Kristan},
  url = {http://arxiv.org/abs/1701.08213},
  journal={arXiv:1701.08213},
  eprint={1701.08213},
  year={2017}
}

@article{Lykos1963,
  title = {Discussion on The {H}artree-{F}ock Approximation},
  author = {Lykos, P. and Pratt, G. W.},
  journal = {Rev. Mod. Phys.},
  volume = {35},
  issue = {3},
  pages = {496--501},
  numpages = {0},
  year = {1963},
  month = {Jul},
  publisher = {American Physical Society},
  doi = {10.1103/RevModPhys.35.496},
  url = {https://link.aps.org/doi/10.1103/RevModPhys.35.496}
}

@article{iQCC,
author = {Ryabinkin, Ilya G. and Lang, Robert A. and Genin, Scott N. and Izmaylov, Artur F.},
title = {Iterative Qubit Coupled Cluster Approach with Efficient Screening of Generators},
journal = {J. Chem. Theory Comput.},
volume = {16},
number = {2},
pages = {1055-1063},
year = {2020}
}

@article{iQCC_ILC,
author = {Lang, Robert A. and Ryabinkin, Ilya G. and Izmaylov, Artur F.},
title = {Unitary Transformation of the Electronic {H}amiltonian with an Exact Quadratic Truncation of the {B}aker-{C}ampbell-{H}ausdorff Expansion},
journal = {J. Chem. Theory Comput.},
volume = {17},
number = {1},
pages = {66-78},
year = {2021}
}

@article{openfermion,
year = {2020},
publisher = {IOP Publishing},
volume = {5},
number = {3},
pages = {034014},
author = {Jarrod R McClean and Nicholas C Rubin and Kevin J Sung and Ian D Kivlichan and Xavier Bonet-Monroig and Yudong Cao and Chengyu Dai and E Schuyler Fried and Craig Gidney and Brendan Gimby and Pranav Gokhale and Thomas Häner and Tarini Hardikar and Vojtěch Havlíček and Oscar Higgott and Cupjin Huang and Josh Izaac and Zhang Jiang and Xinle Liu and Sam McArdle and Matthew Neeley and Thomas O’Brien and Bryan O’Gorman and Isil Ozfidan and Maxwell D Radin and Jhonathan Romero and Nicolas P D Sawaya and Bruno Senjean and Kanav Setia and Sukin Sim and Damian S Steiger and Mark Steudtner and Qiming Sun and Wei Sun and Daochen Wang and Fang Zhang and Ryan Babbush},
title = {OpenFermion: the electronic structure package for quantum computers},
journal = {Quantum Sci. Technol.}
}

@book{Helgaker,
address = {Chichester, UK},
author = {Helgaker, Trygve and J{\o}rgensen, Poul and Olsen, Jeppe},
publisher = {John Wiley \& Sons, Ltd},
title = {{Molecular Electronic-Structure Theory}},
year = {2000}
}

@article{pyscf,
    author = {Sun, Qiming and Zhang, Xing and Banerjee, Samragni and Bao, Peng and Barbry, Marc and Blunt, Nick S. and Bogdanov, Nikolay A. and Booth, George H. and Chen, Jia and Cui, Zhi-Hao and Eriksen, Janus J. and Gao, Yang and Guo, Sheng and Hermann, Jan and Hermes, Matthew R. and Koh, Kevin and Koval, Peter and Lehtola, Susi and Li, Zhendong and Liu, Junzi and Mardirossian, Narbe and McClain, James D. and Motta, Mario and Mussard, Bastien and Pham, Hung Q. and Pulkin, Artem and Purwanto, Wirawan and Robinson, Paul J. and Ronca, Enrico and Sayfutyarova, Elvira R. and Scheurer, Maximilian and Schurkus, Henry F. and Smith, James E. T. and Sun, Chong and Sun, Shi-Ning and Upadhyay, Shiv and Wagner, Lucas K. and Wang, Xiao and White, Alec and Whitfield, James Daniel and Williamson, Mark J. and Wouters, Sebastian and Yang, Jun and Yu, Jason M. and Zhu, Tianyu and Berkelbach, Timothy C. and Sharma, Sandeep and Sokolov, Alexander Yu. and Chan, Garnet Kin-Lic},
    title = "{Recent developments in the PySCF program package}",
    journal = {J. Chem. Phys.},
    volume = {153},
    number = {2},
    pages = {024109},
    year = {2020}
}

@article{Higgott2019,
  title = {Variational {Q}uantum {C}omputation of {E}xcited {S}tates},
  author = {Higgott, Oscar and Wang, Daochen and Brierley, Stephen},
  journal = {{Quantum}},
  issn = {2521-327X},
  publisher = {{Verein zur F{\"{o}}rderung des Open Access Publizierens in den Quantenwissenschaften}},
  volume = {3},
  pages = {156},
  month = jul,
  year = {2019}
}

@article{Xie2022,
author = {Xie, Qing-Xing and Liu, Sheng and Zhao, Yan},
title = {Orthogonal State Reduction Variational Eigensolver for the Excited-State Calculations on Quantum Computers},
journal = {Journal of Chemical Theory and Computation},
volume = {18},
number = {6},
pages = {3737-3746},
year = {2022}
}

@article{gvp,
  title = {Rayleigh-Ritz variational principle for ensembles of fractionally occupied states},
  author = {Gross, E. K. U. and Oliveira, L. N. and Kohn, W.},
  journal = {Phys. Rev. A},
  volume = {37},
  issue = {8},
  pages = {2805--2808},
  numpages = {0},
  year = {1988},
  month = {Apr},
  publisher = {American Physical Society},
  doi = {10.1103/PhysRevA.37.2805},
  url = {https://link.aps.org/doi/10.1103/PhysRevA.37.2805}
}

@article{WenYv2021,
author = {Wen, Jingwei and Lv, Dingshun and Yung, Man-Hong and Long, Gui-Lu},
title = {Variational quantum packaged deflation for arbitrary excited states},
journal = {Quantum Eng.},
volume = {3},
number = {4},
pages = {e80},
year = {2021}
}

@article{Holmes2016,
author = {Holmes, Adam A. and Tubman, Norm M. and Umrigar, C. J.},
title = {Heat-Bath Configuration Interaction: An Efficient Selected Configuration Interaction Algorithm Inspired by Heat-Bath Sampling},
journal = {J. Chem. Theory Comput.},
volume = {12},
number = {8},
pages = {3674-3680},
year = {2016}
}

@article{Sharma2017,
author = {Sharma, Sandeep and Holmes, Adam A. and Jeanmairet, Guillaume and Alavi, Ali and Umrigar, C. J.},
title = {Semistochastic Heat-Bath Configuration Interaction Method: Selected Configuration Interaction with Semistochastic Perturbation Theory},
journal = {J. Chem. Theory Comput.},
volume = {13},
number = {4},
pages = {1595-1604},
year = {2017}
}

@ARTICLE{Gratsea2024,
       author = {{Gratsea}, Katerina and {Kottmann}, Jakob S. and {Johnson}, Peter D. and {Kunitsa}, Alexander A.},
        title = "{Comparing Classical and Quantum Ground State Preparation Heuristics}",
      journal = {arXiv e-prints},
         year = 2024,
        month = jan,
          eid = {arXiv:2401.05306},
        pages = {arXiv:2401.05306},
          doi = {10.48550/arXiv.2401.05306},
archivePrefix = {arXiv},
       eprint = {2401.05306},
 primaryClass = {quant-ph},
       adsurl = {https://ui.adsabs.harvard.edu/abs/2024arXiv240105306G}
}

@Article{Peruzzo2014,
author={Peruzzo, Alberto
and McClean, Jarrod
and Shadbolt, Peter
and Yung, Man-Hong
and Zhou, Xiao-Qi
and Love, Peter J.
and Aspuru-Guzik, Al{\'a}n
and O'Brien, Jeremy L.},
title={A variational eigenvalue solver on a photonic quantum processor},
journal={Nature Communications},
year={2014},
month={Jul},
day={23},
volume={5},
number={1},
pages={4213},
issn={2041-1723},
doi={10.1038/ncomms5213},
url={https://doi.org/10.1038/ncomms5213}
}

@article{TILLY20221,
title = {The Variational Quantum Eigensolver: A review of methods and best practices},
journal = {Physics Reports},
volume = {986},
pages = {1-128},
year = {2022},
note = {The Variational Quantum Eigensolver: a review of methods and best practices},
issn = {0370-1573},
doi = {https://doi.org/10.1016/j.physrep.2022.08.003},
url = {https://www.sciencedirect.com/science/article/pii/S0370157322003118},
author = {Jules Tilly and Hongxiang Chen and Shuxiang Cao and Dario Picozzi and Kanav Setia and Ying Li and Edward Grant and Leonard Wossnig and Ivan Rungger and George H. Booth and Jonathan Tennyson}
}

@article{McClean_2016vqe,
doi = {10.1088/1367-2630/18/2/023023},
url = {https://dx.doi.org/10.1088/1367-2630/18/2/023023},
year = {2016},
month = {feb},
publisher = {IOP Publishing},
volume = {18},
number = {2},
pages = {023023},
author = {Jarrod R McClean and Jonathan Romero and Ryan Babbush and Alán Aspuru-Guzik},
title = {The theory of variational hybrid quantum-classical algorithms},
journal = {New Journal of Physics},
}

@Article{Cerezo2021,
author={Cerezo, M.
and Arrasmith, Andrew
and Babbush, Ryan
and Benjamin, Simon C.
and Endo, Suguru
and Fujii, Keisuke
and McClean, Jarrod R.
and Mitarai, Kosuke
and Yuan, Xiao
and Cincio, Lukasz
and Coles, Patrick J.},
title={Variational quantum algorithms},
journal={Nature Reviews Physics},
year={2021},
month={Sep},
day={01},
volume={3},
number={9},
pages={625-644}
}

@article{qAdapt,
  title = {Qubit-ADAPT-VQE: An Adaptive Algorithm for Constructing Hardware-Efficient Ans\"atze on a Quantum Processor},
  author = {Tang, Ho Lun and Shkolnikov, V.O. and Barron, George S. and Grimsley, Harper R. and Mayhall, Nicholas J. and Barnes, Edwin and Economou, Sophia E.},
  journal = {PRX Quantum},
  volume = {2},
  issue = {2},
  pages = {020310},
  numpages = {16},
  year = {2021},
  publisher = {American Physical Society},
}

@article{ADAPT,
author = {Grimsley, Harper R. and Economou, Sophia E. and Barnes, Edwin and Mayhall, Nicholas J.},
journal = {Nat. Commun.},
publisher = {Springer US},
title = {{An adaptive variational algorithm for exact molecular simulations on a quantum computer}},
volume = {10},
pages = {3007},
year = {2019}
}

@Article{Mills2000,
author={Mills, Patrick
and Tsang, Edward},
title={Guided Local Search for Solving {SAT} and Weighted {MAX-SAT} Problems},
journal={Journal of Automated Reasoning},
year={2000},
month={Feb},
day={01},
volume={24},
number={1},
pages={205-223},
}

@article{Genin2022,
author = {Genin, Scott N. and Ryabinkin, Ilya G. and Paisley, Nathan R. and Whelan, Sarah O. and Helander, Michael G. and Hudson, Zachary M.},
title = {Estimating Phosphorescent Emission Energies in IrIII Complexes Using Large-Scale Quantum Computing Simulations**},
journal = {Angewandte Chemie International Edition},
volume = {61},
number = {19},
pages = {e202116175},
year = {2022}
}

@ARTICLE{Domcke2011,
       author = {{Domcke}, Wolfgang and {Yarkony}, David R. and {K{\"o}ppel}, Horst},
        title = "{Conical Intersections}",
      journal = {Advanced Series in Physical Chemistry},
         year = 2011,
       volume = {17},
}

@article{Yarkony2012,
author = {Yarkony, David R.},
title = {Nonadiabatic Quantum Chemistry—Past, Present, and Future},
journal = {Chem. Rev.},
volume = {112},
number = {1},
pages = {481-498},
year = {2012}
}

@article{Aspuru2005,
author = {Alán Aspuru-Guzik  and Anthony D. Dutoi  and Peter J. Love  and Martin Head-Gordon },
title = {Simulated Quantum Computation of Molecular Energies},
journal = {Science},
volume = {309},
number = {5741},
pages = {1704-1707},
year = {2005}
}

@article{Griffiths1996,
  title = {Semiclassical Fourier Transform for Quantum Computation},
  author = {Griffiths, Robert B. and Niu, Chi-Sheng},
  journal = {Phys. Rev. Lett.},
  volume = {76},
  issue = {17},
  pages = {3228--3231},
  numpages = {0},
  year = {1996},
  month = {Apr},
  publisher = {American Physical Society},
  doi = {10.1103/PhysRevLett.76.3228},
  url = {https://link.aps.org/doi/10.1103/PhysRevLett.76.3228}
}

@article{Lin_Tong_2022,
  title = {Heisenberg-Limited Ground-State Energy Estimation for Early Fault-Tolerant Quantum Computers},
  author = {Lin, Lin and Tong, Yu},
  journal = {PRX Quantum},
  volume = {3},
  issue = {1},
  pages = {010318},
  numpages = {21},
  year = {2022},
  month = {Feb},
  publisher = {American Physical Society},
  doi = {10.1103/PRXQuantum.3.010318},
  url = {https://link.aps.org/doi/10.1103/PRXQuantum.3.010318}
}

@article{Wang_Franca_2023,
  title = {Quantum algorithm for ground state energy estimation using circuit depth with exponentially improved dependence on precision},
  author = {Wang, Guoming and Fran{\c{c}}a, Daniel Stilck and Zhang, Ruizhe and Zhu, Shuchen and Johnson, Peter D.},
  journal = {{Quantum}},
  issn = {2521-327X},
  publisher = {{Verein zur F{\"{o}}rderung des Open Access Publizierens in den Quantenwissenschaften}},
  volume = {7},
  pages = {1167},
  year = {2023}
}

@article{Ding_Lin_2023,
  title = {Even Shorter Quantum Circuit for Phase Estimation on Early Fault-Tolerant Quantum Computers with Applications to Ground-State Energy Estimation},
  author = {Ding, Zhiyan and Lin, Lin},
  journal = {PRX Quantum},
  volume = {4},
  issue = {2},
  pages = {020331},
  numpages = {30},
  year = {2023},
  month = {May},
  publisher = {American Physical Society},
  doi = {10.1103/PRXQuantum.4.020331},
  url = {https://link.aps.org/doi/10.1103/PRXQuantum.4.020331}
}

@article{BALKOVA1992,
title = {Coupled-cluster method for open-shell singlet states},
journal = {Chem. Phys. Lett.},
volume = {193},
number = {5},
pages = {364-372},
year = {1992},
issn = {0009-2614},
author = {Anna Balková and Rodney J. Bartlett},
}

@article{ChanSharma2011,
   author = "Chan, Garnet Kin-Lic and Sharma, Sandeep",
   title = "The Density Matrix Renormalization Group in Quantum Chemistry", 
   journal= "Annu Rev Phys Chem",
   year = "2011",
   volume = "62",
   number = "Volume 62, 2011",
   pages = "465-481",
   doi = "https://doi.org/10.1146/annurev-physchem-032210-103338",
   url = "https://www.annualreviews.org/content/journals/10.1146/annurev-physchem-032210-103338",
   publisher = "Annual Reviews",
  }

@article{Chandross1999,
  title = {Density-matrix renormalization-group method for excited states},
  author = {Chandross, M. and Hicks, J. C.},
  journal = {Phys. Rev. B},
  volume = {59},
  issue = {15},
  pages = {9699--9702},
  numpages = {0},
  year = {1999},
  month = {Apr},
  publisher = {American Physical Society},
  doi = {10.1103/PhysRevB.59.9699},
  url = {https://link.aps.org/doi/10.1103/PhysRevB.59.9699}
}

@article{HuChan2015,
author = {Hu, Weifeng and Chan, Garnet Kin-Lic},
title = {Excited-State Geometry Optimization with the Density Matrix Renormalization Group, as Applied to Polyenes},
journal = {J. Chem. Theory Comput.},
volume = {11},
number = {7},
pages = {3000-3009},
year = {2015}
}

@inbook{Freitag2020,
author = {Freitag, Leon and Reiher, Markus},
publisher = {John Wiley \& Sons, Ltd},
isbn = {9781119417774},
title = {The Density Matrix Renormalization Group for Strong Correlation in Ground and Excited States},
booktitle = {Quantum Chemistry and Dynamics of Excited States},
chapter = {7},
pages = {205-245},
year = {2020}
}

@article{Barford2001,
  title = {Density-matrix renormalization-group calculations of excited states of linear polyenes},
  author = {Barford, William and Bursill, Robert J. and Lavrentiev, Mikhail Yu},
  journal = {Phys. Rev. B},
  volume = {63},
  issue = {19},
  pages = {195108},
  numpages = {8},
  year = {2001},
  month = {Apr},
}

@article{Roos1980,
title = {A complete active space SCF method (CASSCF) using a density matrix formulated super-CI approach},
journal = {Chem. Phys.},
volume = {48},
number = {2},
pages = {157-173},
year = {1980},
author = {Björn O. Roos and Peter R. Taylor and Per E.M. Sigbahn},
}

@article{Olsen1988,
    author = {Olsen, Jeppe and Roos, Björn O. and J{\o}rgensen, Poul and Jensen, Hans J{\o}rgen Aa.},
    title = "{Determinant based configuration interaction algorithms for complete and restricted configuration interaction spaces}",
    journal = {J. Chem. Phys.},
    volume = {89},
    number = {4},
    pages = {2185-2192},
    year = {1988}
}

@article{Malmqvist1990,
author = {Malmqvist, Per Aake. and Rendell, Alistair. and Roos, Bjoern O.},
title = {The restricted active space self-consistent-field method, implemented with a split graph unitary group approach},
journal = {J. Phys. Chem.},
volume = {94},
number = {14},
pages = {5477-5482},
year = {1990}
}

@article{LiManni2011,
    author = {Li Manni, Giovanni and Aquilante, Francesco and Gagliardi, Laura},
    title = "{Strong correlation treated via effective hamiltonians and perturbation theory}",
    journal = {J. Chem. Phys.},
    volume = {134},
    number = {3},
    pages = {034114},
    year = {2011}
}

@article{Andersson1990,
author = {Andersson, Kerstin. and Malmqvist, Per Aake. and Roos, Bjoern O. and Sadlej, Andrzej J. and Wolinski, Krzysztof.},
title = {Second-order perturbation theory with a CASSCF reference function},
journal = {J. Phys. Chem.},
volume = {94},
number = {14},
pages = {5483-5488},
year = {1990}
}

@article{Andersson1992,
    author = {Andersson, Kerstin and Malmqvist, Per‐Åke and Roos, Björn O.},
    title = "{Second‐order perturbation theory with a complete active space self‐consistent field reference function}",
    journal = {J. Chem. Phys.},
    volume = {96},
    number = {2},
    pages = {1218-1226},
    year = {1992}
}

@article{RoosAndersson1996,
  title={Multiconfigurational perturbation theory: Applications in electronic spectroscopy},
  author={Roos, Bj{\"o}rn O and Andersson, Kerstin and F{\"u}lscher, Markus P and Malmqvist, Per-{\^a}ke and Serrano-Andr{\'e}s, Luis and Pierloot, Kristin and Merch{\'a}n, Manuela},
  journal={Advances in chemical physics: new methods in computational quantum mechanics},
  volume={93},
  pages={219--331},
  year={1996},
  publisher={John Wiley \& Sons, Inc. Hoboken, NJ, USA}
}

@article{delta_SCF, 
title={On the calculation of multiplet energies by the hartree-fock-slater method}, 
volume={43}, 
ISSN={0040-5744, 1432-2234}, 
url={http://link.springer.com/10.1007/BF00551551}, 
DOI={10.1007/BF00551551}, 
number={3}, 
journal={Theoretica Chimica Acta}, 
author={Ziegler, Tom and Rauk, Arvi and Baerends, Evert J.}, 
year={1977}, 
pages={261–271} 
}

@article{delta_ADAPT_VQE,
author = {Nykänen, Anton and Thiessen, Leander and Borrelli, Elsi-Mari and Krishna, Vijay and Knecht, Stefan and Pavošević, Fabijan},
title = {Toward Accurate Calculation of Excitation Energies on Quantum Computers with $\Delta$ADAPT-VQE: A Case Study of BODIPY Derivatives},
journal = {The Journal of Physical Chemistry Letters},
volume = {15},
number = {28},
pages = {7111-7117},
year = {2024},
doi = {10.1021/acs.jpclett.4c01301},
note = {PMID: 38954795},
URL = {https://doi.org/10.1021/acs.jpclett.4c01301}
}

@article{ADAPT_VQE_SCF,
author = {Fitzpatrick, Aaron and Nykänen, Anton and Talarico, N. Walter and Lunghi, Alessandro and Maniscalco, Sabrina and García-P{\'e}rez, Guillermo and Knecht, Stefan},
title = {Self-Consistent Field Approach for the Variational Quantum Eigensolver: Orbital Optimization Goes Adaptive},
journal = {The Journal of Physical Chemistry A},
volume = {128},
number = {14},
pages = {2843-2856},
year = {2024},
doi = {10.1021/acs.jpca.3c05882},
URL = {https://doi.org/10.1021/acs.jpca.3c05882}
}

@article{Grimsley_2025,
doi = {10.1088/2058-9565/ad9fa2},
url = {https://doi.org/10.1088/2058-9565/ad9fa2},
year = {2025},
month = {jan},
publisher = {IOP Publishing},
volume = {10},
number = {2},
pages = {025003},
author = {Grimsley, Harper R and Evangelista, Francesco A},
title = {Challenging excited states from adaptive quantum eigensolvers: subspace expansions vs. state-averaged strategies},
journal = {Quantum Science and Technology}
}

@article{SPD,
author = {Tomislav Begušić  and Johnnie Gray  and Garnet Kin-Lic Chan },
title = {Fast and converged classical simulations of evidence for the utility of quantum computing before fault tolerance},
journal = {Science Advances},
volume = {10},
number = {3},
pages = {eadk4321},
year = {2024},
doi = {10.1126/sciadv.adk4321},
URL = {https://www.science.org/doi/abs/10.1126/sciadv.adk4321}
}

@article{Delta_SCF_MOM,
author = {Gilbert, Andrew T. B. and Besley, Nicholas A. and Gill, Peter M. W.},
title = {Self-Consistent Field Calculations of Excited States Using the Maximum Overlap Method (MOM)},
journal = {The Journal of Physical Chemistry A},
volume = {112},
number = {50},
pages = {13164-13171},
year = {2008},
doi = {10.1021/jp801738f},
    note ={PMID: 18729344},
URL = {https://doi.org/10.1021/jp801738f}
}

@misc{200iQCC,
Author = {Scott N. Genin and Ohyun Kwon and Seyyed Mehdi Hosseini Jenab and Seon-Jeong Lim and Taehyung Kim and Tae-Gon Kim and Rami Gherib and Angela F. Harper and Ilya G. Ryabinkin and Michael G. Helander},
Title = {Towards Quantum Advantage in Chemistry},
Year = {2025},
Eprint = {arXiv:2512.13657},
}

@misc{majorana_products,
Author = {Aaron Miller and Joachim Favre and Zoë Holmes and Özlem Salehi and Rahul Chakraborty and Anton Nykänen and Zoltán Zimborás and Adam Glos and Guillermo García-Pérez},
Title = {Simulation of Fermionic circuits using Majorana Propagation},
Year = {2025},
Eprint = {arXiv:2503.18939},
}

\appendix 

\clearpage

\section{DIS construction for  multiconfigurational reference states} \label{app:dis_mr}

Herein, the DIS construction for an ensemble of multiconfigurational reference states is provided. In Section \ref{sec:ms_dis}, it is shown that the ensemble-averaged energy gradient for candidate generator $\hat T_{\alpha}$ has the form Eq. (\ref{eq:g_sa}) for ensembles over determinantal references, i.e., $\hat \rho = \sum_{i} \ket{\phi_i} \bra{\phi_i}/N_s$. In the multiconfigurational case, we utilize the more general ensemble $\rho = \sum_{I}  \ket{I} \bra{I} /N_s$, with 
\begin{align} \label{eq:mc_reference}
\ket{I} = \sum_{i} c^{(I)}_i \ket{\phi_i}.
\end{align}
It is shown here that we can obtain a form of the multiconfigurational state-averaged gradient:
\begin{align} \label{eq:g_sa_mc}
g_{\alpha} = \dfrac{1}{N_s}\left| \sum_{I=1}^{N_s} \mathrm{Im} \left( \bra{I} \hat H \hat T_{\alpha} \ket{I}  \right)   \right|
\end{align}
which retains the general form of Eq. (\ref{eq:g_sa}), but with modifications to $\Xi^{(\alpha)}$ components. Firstly, let $\hat \Omega_i^{(I)}$ be the \textit{wave operator} which achieves $\hat \Omega_i^{(I)} \ket{\phi_i} = \ket{I}$. From Eq. (\ref{eq:mc_reference}), such operator can be written in the form
\begin{align}
\hat \Omega_i^{(I)} = \sum_{j} c_j^{(I)} \hat X_{ij},
\end{align}
where $\ket{\phi_j} = \hat X_{ij} \ket{\phi_i}$, hence $\hat X_{ii} = \hat 1$. Note that $\hat \Omega_i^{(I)}$ is real-valued when $\ket{I}$ is real, as assumed here. In Eq. (\ref{eq:g_sa_mc}), we can explicitly expand $\ket{I}$, and substitute $\bra{I}$ with $\bra{\phi_i} \hat \Omega_i^{(I)}$ to obtain
\begin{align}
g_{\alpha} & = \dfrac{1}{N_s}\left| \sum_{I=1}^{N_s} \sum_{i=1}^{L}  c_i^{(I)}\mathrm{Im} \left( \bra{\phi_i} \hat \Omega_{i}^{(I)} \hat H \hat T_{\alpha} \ket{\phi_i}  \right)  \right| \nonumber \\ 
& = \dfrac{1}{N_s}\left| \mathrm{Im} \left( \sum_{i=1}^{L} \bra{\phi_i} \hat H^{(SA)}_i \hat T_{\alpha} \ket{\phi_i}    \right)   \right| \label{eq:gradient_app}
\end{align}
where we have defined 
\begin{align} \label{eq:H_sa}
\hat H^{(SA)}_i = \sum_{I=1}^{N_s} c_i^{(I)} \hat \Omega_i^{(I)} \hat H,
\end{align}
which is a linear combination of $\hat H$ left-multiplied by $\hat X_{ij}$ operators. $L$ is the total number of unique Slater determinants used across all multireference states $\ket{I}$. Factorizing candidate Pauli term $\hat T_{\alpha}$ as $\theta_{\alpha} \hat X_{\alpha} \hat Z_{\alpha}$ leads to 
\begin{align}
g_{\alpha} = \dfrac{\left| \mathrm{Im} (\theta_{\alpha}) \right| }{N_s} \left| \sum_{i=1}^{L} \lambda_{i}^{(\alpha)} \Xi_i^{(\alpha)}  \right|, \label{eq:g_sa_h_sa}
\end{align}
with $\lambda^{(\alpha)}_i = \bra{\phi_i}  \hat  Z_{\alpha} \ket{\phi_i}$, and $\Xi_i^{(\alpha)} = \bra{\phi_i} \hat H_{i}^{(SA)} \hat X_{\alpha} \ket{\phi_i}$. From Eq. (\ref{eq:g_sa_h_sa}), the multiconfigurational ensemble energy gradient exhibits the same decoupling of the $\hat X_{\alpha}$ and $\hat Z_{\alpha}$ degrees of freedom in $\hat T_{\alpha}$, up to $\theta_{\alpha}$, as for the case of ensembles of single determinant references in Section \ref{sec:ms_dis}. The $\hat Z_{\alpha}$ choice fully dictates the relative phases $\{ \lambda_i^{(\alpha)} \}_i$, whereas $\hat X_{\alpha}$'s role is in expectation values $\Xi_i^{(\alpha)}$. Such expectation values can be connected to expectation values of the generalized Ising parts in $\hat H$ in Eq. (\ref{eq:H_ising}) following simple rules, which we now derive. 

By expanding $\hat \Omega_i^{(I)}$ in Eq. (\ref{eq:H_sa}), and inserting Ising factorized form of $\hat H = \sum_k \hat D_k \hat X_k = \sum_k \hat X_k \hat D_k$ (with second equality holding due to assumed realness of $\hat H$), we obtain 
\begin{align}
\hat H_{i}^{(SA)} & = \sum_{I=1}^{N_s} c_i^{(I)} \sum_{j=1}^{L} \sum_k c_j^{(I)} \hat X_{ij}  \hat X_k \hat D_k \\ 
& = \sum_{I=1}^{N_s} c_i^{(I)} \sum_{j=1}^{L} \sum_k c_j^{(I)} \hat X_k^{(ij)} \hat D_k,
\end{align}
where we have defined $\hat X_{k}^{(ij)} = \hat X_{ij} \hat X_k$. This leads to
\begin{align} \label{eq:h_sa_exp}
\Xi_i^{(\alpha)} = \sum_{I=1}^{N_s}  \sum_{j=1}^{L} \sum_k c_i^{(I)} c_j^{(I)}  \bra{\phi_i} \hat X_k^{(ij)} \hat X_{\alpha} \ket{\phi_i} \nonumber \\
\times \bra*{\hat X_k^{(ij)} \phi_i} \hat D_k \ket*{\hat X_k^{(ij)} \phi_i},
\end{align}
where $\ket*{\hat X_k^{(ij)} \phi_i} = \hat X_k^{(ij)}\ket{\phi_i}$. Hence, $\Xi_i^{(\alpha)}$ is zero unless there exists at least one instance of $\hat X_{k}^{(ij)} = \hat X_{\alpha}$, which additively contributes $w_I c_i^{(I)} c_j^{(I)} \bra*{\hat X_k^{(ij)} \phi_i} \hat D_k \ket*{\hat X_k^{(ij)} \phi_i}$ to $\Xi^{(\alpha)}_i$. To efficiently screen the $\hat X_{\alpha}$'s which lead to non-zero $g_{\alpha}$, we consider all possible $\hat X_k^{(ij)}$'s, i.e., possible products from coupling $\hat X_{ij}$ operators in the wave operators $\hat \Omega_i^{(I)}$'s, and $\hat X_k$'s appearing in the Ising factorized form of $\hat H$, Eq. (\ref{eq:H_ising}).

\section{Phase-alignment procedures} \label{app:phase_align}

For general ensembles of reference states, the energy gradient magnitude expression for generator $\hat T_\alpha$ takes the form of
\begin{align} \label{eq:truegrad_app}
g_{\alpha} = \dfrac{| \mathrm{Im}(\theta_\alpha) |}{N_s} \left| \sum_{i=1}^{L} \lambda_i^{(\alpha)} \Xi^{(\alpha)}_i \right|,
\end{align}
Recall that for a given $\hat T_\alpha$ in the direct interaction space, $\lambda_i^{(\alpha)}$ depends only on the choice of $\hat{Z}_\alpha$, and $\Xi_i^{(\alpha)}$ depends only on $\hat{X}_\alpha$. We represent the operators $\hat{Z}_\alpha$ and $\hat{X}_\alpha$ as tensor products
\begin{align}
\hat Z_\alpha = \prod_{p=1}^{N_q} \hat z_p^{\nu_p^{(\alpha)}} \label{eq:Z_tenosr} \\
\hat X_\alpha = \prod_{p=1}^{N_q} \hat x_p^{\mu_p^{(\alpha)}} \label{eq:X_tenosr}
\end{align}
where $\mu^{(\alpha)}_p,\ \nu^{(\alpha)}_p \in \{0, 1\}$ are the $p^{\text{th}}$ elements of the vectors $\vec{\mu}^{(\alpha)}$ and $\vec{\nu}^{(\alpha)}$ respectively. In this Appendix, we describe how to select $\hat Z_{\alpha}$ which maximizes $g_{\alpha}$, for a given $\hat X_{\alpha}$. We formulate this problem in the domain of $\hat Z_{\alpha}$'s binary representation, $\vec{\nu}^{(\alpha)}$. We describe the strategy for finding optimal $\vec{\nu}^{(\alpha)}$ in Appendix \ref{sec:selection_opt}, and a heuristic yet efficient strategy in Appendix \ref{sec:selection_greedysat}.

Recall $g_\alpha = 0$ unless $\theta_\alpha \in \{ i, -i \}$, which is the case when $\hat X_{\alpha}$ and $\hat Z_{\alpha}$ have odd overlapping support. In terms of the binary vectors, this leads to the requirement 
\begin{align} \label{eq: mu_nu_constraint}
\vec{\mu}^{(\alpha)} \cdot \vec{\nu}^{(\alpha)} \mod 2 = 1.
\end{align}
The relative phases $\lambda_i^{(\alpha)}$ are obtained as,
\begin{align} \label{eq:lambda_binary}
\lambda_i^{(\alpha)} = \bra{\phi_i} \hat Z_{\alpha} \ket{\phi_i} = \prod_{p=1}^{N_q} \bra*{\phi_i^{(p)}} \hat z_{\alpha}^{\nu^{(\alpha)}_p} \ket*{\phi_i^{(p)}} \in \{1, -1 \},
\end{align}
where $\ket*{\phi_i^{(p)}} \in \{ \ket{0}, \ket{1} \}$. Let us introduce an $N_q$-bit binary vector $\vec{\phi}^{(i)} = (\phi^{(i)}_1, \hdots \phi^{(i)}_{N_q})$ representing the computational basis state $\ket{\phi_i}$, that is 
\begin{align}
\vec{\phi}^{(i)}_p = \begin{cases} 1 \text{ if } \ket{\phi_i^{(p)}} = \ket{1} \\ 0 \text{ if } \ket{\phi_i^{(p)}} = \ket{0}  \end{cases}.
\end{align}
Phase $\lambda_{i}^{(\alpha)} = -1$ if and only if there are an odd number of instances where $\nu_p^{(\alpha)} = \phi_p^{(i)} = 1$, otherwise $\lambda_{i}^{(\alpha)} = 1$. Hence, $\lambda_{i}^{(\alpha)}$ is written as a function of binary vectors $\vec{\nu}^{(\alpha)}$ and $\vec{\phi}^{(i)}$ as 
\begin{align} \label{eq:lambdas_as_vecfuncs}
\lambda_i^{(\alpha)} =  1 - 2 \left( \vec{\phi}^{(i)} \cdot \vec{\nu}^{(\alpha)} \mod 2 \right).
\end{align}
Thus we can express the gradient (up to normalization by $N_s$) as a cost function in terms of $\nu^{(\alpha)}$ and formulate the problem as an $N_q$ bit constrained binary optimization
\begin{align}
\begin{split} \label{eq:phase_align_formal}
    & \max_{\vec{\nu}^{(\alpha)}} C(\vec{\nu}^{(\alpha)})  \\ 
    & \text{  subject to: }  \vec{\mu}^{(\alpha)} \cdot \vec{\nu}^{(\alpha)} \mod 2 = 1
\end{split}
\end{align}
where 
\begin{align} \label{eq:opt_cost}
C(\vec{\nu}^{(\alpha)}) = \left| \sum_{i=1}^{L} \Xi_i^{(\alpha)} \left[ 1 - 2 \left( \vec{\phi}^{(i)} \cdot \vec{\nu}^{(\alpha)} \mod 2 \right) \right] \right|.
\end{align}
We will look at two different strategies at solving this binary optimization problem.\\

\subsection{Optimal strategy} \label{sec:selection_opt}
To find optimal solution to \eqref{eq:phase_align_formal}, note that the cost function, Eq. (\ref{eq:opt_cost}), consists of $L$ distinct $N_q$-bit clauses, and hence is generally exponentially hard to find the optimal value of $\vec{\nu}^{(\alpha)}$. Since $\vec{\Xi}_i^{(\alpha)}$ has been precomputed, evaluation of Eq. (\ref{eq:opt_cost}) is computationally fast, with scaling $O(LN_q)$. We find the optimal solution by brute force search of the constrained space of binary vectors. Since this is implausible for sufficiently large $N_q$, heuristic binary optimization strategies can be employed. In the next section, we discuss one such heuristic method. 

\subsection{Greedy satisfiability} \label{sec:selection_greedysat}

It turns out one can formulate the question, is there a $\vec{\nu}^{(\alpha)}$ which aligns signs of \textit{all} terms $\{ \lambda_i^{(\alpha)} \Xi_i^{(\alpha)}  \}_{i=1}^{L}$ in Eq. (\ref{eq:truegrad_app}), as an efficiently solvable satisfiability problem. If satisfied, the algorithm returns the satisfying $\vec{\nu}^{(\alpha)}$. If unsatisfiable, one can remove constraints existing in the problem and check again for satisfiability. We refer to this procedure as the \texttt{GreedySAT} phase-alignment routine, and explain it in further detail below.

The idealized scenario occurs when for all nonzero $\Xi_i^{(\alpha)}$,
\begin{align} \label{eq:parity_alignment}
\text{sgn}(\lambda_i^{(\alpha)} \Xi_i^{(\alpha)}) = \text{sgn}(\lambda_l^{(\alpha)} \Xi_l^{(\alpha)}),
\end{align}
where sgn$(x)$ is the signum function for $x \in \mathbb{R}$,
\begin{align}
\text{sgn}(x) = \begin{cases} 1 \text{ if } x > 0 \\ 0 \text{ if } x = 0 \\ -1 \text{ if } x < 0. \end{cases}
\end{align}
To satisfy Eq. (\ref{eq:parity_alignment}), two equivalent assignments exist,
\begin{align} \label{eq:2nd_cond}
\lambda^{(\alpha)}_i = (\pm) \text{sgn}\left(\Xi_i^{(\alpha)}\right),
\end{align}
for all considered $\Xi_i^{(\alpha)}$, where the choice of $(\pm)$ is held fixed for all $i$.

Attempting to find $\vec{\nu}^{(\alpha)}$ which satisfies Eq. (\ref{eq:2nd_cond}) for all $L' \leq L$ non-zero $\Xi_i^{(\alpha)}$'s, along with the requirement of Eq. \eqref{eq: mu_nu_constraint}, leads to a system of $L' + 1$ equations on the $N_q$ binary variables in $\vec{\nu}^{(\alpha)}$. To see this, substitute Eq. \eqref{eq:2nd_cond} in Eq. \eqref{eq:lambdas_as_vecfuncs} to get
\begin{align} \label{eq:2nd_cond_binary}
\vec{\phi}^{(i)} \cdot \vec{\nu}^{(\alpha)}  \mod 2 = \frac{1 - (\pm)\text{sgn}(\Xi_i^{(\alpha)})}{2}
\end{align}
which are $L'$ set of equations for $i\in\{1\dots L'\}$, and one more equation comes from the constraint in Eq. \eqref{eq: mu_nu_constraint}. We can then formulate the system of equations as 
\begin{align} \label{eq:system_gf2}
\boldsymbol{M} \vec{\nu}^{(\alpha)} = \vec{\boldsymbol{b}}\ \ \ \text{over} \ \mathbb{F}_2,
\end{align}
where $\boldsymbol{M}$ is a $(L'+1) \times N_q$ matrix,
\begin{align}
\boldsymbol{M} = 
\begin{pmatrix}    
\vec{\mu}^{(\alpha)}_1 & \hdots & \vec{\mu}^{(\alpha)}_{N_q}  \\ 
\vec{\phi}^{(1)}_1 & \hdots & \vec{\phi}^{(1)}_{N_q} \\
& \vdots & \\ 
\vec{\phi}^{(L')}_1 & \hdots & \vec{\phi}^{(L')}_{N_q}
\end{pmatrix},
\end{align}
and $\vec{\boldsymbol{b}}$ is a $L'+1$ dimensional vector, 
\begin{align} \label{eq:vec_b_mat}
\vec{\boldsymbol{b}} = 
\begin{pmatrix}
1 \\ 
\left[1- (\pm)\text{sgn}(\Xi_1^{(\alpha)}) \right] /2 \\ 
\vdots \\
\left[1- (\pm)\text{sgn}(\Xi_{L'}^{(\alpha)}) \right] /2
\end{pmatrix}.
\end{align}
The restriction of matrix arithmetic to the space $\mathbb{F}_2$ effectively ensures the \textit{modulo} 2 operations in equations \eqref{eq: mu_nu_constraint} and \eqref{eq:2nd_cond_binary}. We can solve Eq. (\ref{eq:system_gf2}) by binary Gaussian elimination, for instance. In our numerical examples, we utilize the \texttt{SageMath} package to this end. 

We now describe the complete \texttt{GreedySAT} procedure below, which includes a prescription for when no solution to Eq. (\ref{eq:system_gf2}) can be found. Essentially, if no solution can be found for considering the phase-alignment of all $L'$ terms via satisfying all Eq. (\ref{eq:2nd_cond_binary}), we remove consideration of a specific instance of Eq. (\ref{eq:2nd_cond_binary}) associated with the lowest valued $| \Xi_{k}^{(\alpha)} |$, and attempt to solve the system of fewer equations. This removal of least-important constraints is iteratively performed until a solution $\vec{\nu}^{(\alpha)}$ has been found, and the true value of the associated $g_\alpha$ is computed via Eq. (\ref{eq:opt_cost}). The procedure can be summarized as follows:

\begin{enumerate}
\item For a given $\vec\mu^{(\alpha)}$ corresponding to an element of the direct interaction space, evaluate $\vec{\boldsymbol{\Xi}}^{(\alpha)} = (\Xi_1^{(\alpha)}, \Xi_2^{(\alpha)}\dots\Xi_{L'}^{(\alpha)})$, where $\vec{\boldsymbol{\Xi}}^{(\alpha)} $ has only non zero elements, arranged in decreasing order of their absolute values.

\item  Attempt to find solution to Eq. (\ref{eq:system_gf2}) for the $(+)$ assignment in Eq. (\ref{eq:vec_b_mat}). If no solution exists, attempt to solve the same system with the $(-)$ assignment. If a solution has been found for either $(\pm)$ assignments, go to Step 4 with solution $\vec{\nu}^{(\alpha)}$ to Eq. (\ref{eq:system_gf2}), otherwise, enter Step 3.

\item Redefine $\vec{\boldsymbol{\Xi}}^{(\alpha)} = (\Xi_1^{(\alpha)}, \Xi_2^{(\alpha)}\dots\Xi_{L'}^{(\alpha)})$ with $L' \to L' - 1$, by dropping the smallest absolute element of $\vec{\boldsymbol{\Xi}}^{(\alpha)}$ and re-enter step 2.

\item Once a solution $\vec{\nu}^{(\alpha)}$ has been found, the corresponding Pauli term $\hat T_\alpha$ has $\hat X_{\alpha}$ and $\hat Z_{\alpha}$ parts given by $\vec{\mu}^{(\alpha)}$ and $\vec{\nu}^{(\alpha)}$ respectively, and its gradient magnitude $g_{\alpha}$ is given by inserting the found $\vec{\nu}^{(\alpha)}$ into Eq. (\ref{eq:opt_cost}). This resulting $\hat T_\alpha$ represents the highest $g_\alpha$ candidate found by the \texttt{GreedySAT} routine.
\end{enumerate}

\subsection{Numerical assessment of \texttt{GreedySAT}}

\begin{table*}[!htb]
\caption{Empirical average gradient ratios $R_{avg}$ for the \texttt{GreedySAT} routine applied to uniformly random phase-alignment problems of $N_q$ qubits and $L$-dimensional model space, obtained using $100$ samples of problems. One standard deviation $\sigma$ is included as $\pm \sigma$ . }
\centering
\scriptsize

\begin{tabular}{|c|cccccccccc|}
\hline
                        & \multicolumn{10}{c|}{$L$}                                                                                                                                                                                                                                                                                                                                                                                                                                                                                                                                                                                                                                                                                                                                                                                                                                                                                                                                                                                                                                                                                                                                                                                                                                                                                                                                                                                                                                   \\ \cline{2-11} 
\multirow{-2}{*}{$N_q$} & \multicolumn{1}{c|}{2}                                                                                                                                        & \multicolumn{1}{c|}{4}                                                                                                                    & \multicolumn{1}{c|}{6}                                                                                                                    & \multicolumn{1}{c|}{8}                                                                                                                     & \multicolumn{1}{c|}{10}                                                                                                                    & \multicolumn{1}{c|}{12}                                                                                                                    & \multicolumn{1}{c|}{14}                                                                                                                    & \multicolumn{1}{c|}{16}                                                                                                                  & \multicolumn{1}{c|}{18}                                                                                                                    & 20                                                                                                                    \\ \hline
2                       & \multicolumn{1}{c|}{{ \begin{tabular}[c]{@{}c@{}} $1.0 \pm 0.0$\end{tabular}}} &                                                                                                                                           &                                                                                                                                           &                                                                                                                                            &                                                                                                                                            &                                                                                                                                            &                                                                                                                                            &                                                                                                                                          &                                                                                                                                            &                                                                                                                       \\ \cline{1-3}
4                       & \multicolumn{1}{c|}{\begin{tabular}[c]{@{}c@{}} $1.0 \pm 0.0$\end{tabular}}                         & \multicolumn{1}{c|}{\begin{tabular}[c]{@{}c@{}} $0.974 \pm 0.139$\end{tabular}}      &                                                                                                                                           &                                                                                                                                            &                                                                                                                                            &                                                                                                                                            &                                                                                                                                            &                                                                                                                                          &                                                                                                                                            &                                                                                                                       \\ \cline{1-4}
6                       & \multicolumn{1}{c|}{\begin{tabular}[c]{@{}c@{}} $1.0 \pm 0.0$\end{tabular}}                            & \multicolumn{1}{c|}{\begin{tabular}[c]{@{}c@{}} $0.99 \pm 0.064$\end{tabular}} & \multicolumn{1}{c|}{\begin{tabular}[c]{@{}c@{}} $0.975 \pm 0.111$\end{tabular}}  &                                                                                                                                            &                                                                                                                                            &                                                                                                                                            &                                                                                                                                            &                                                                                                                                          &                                                                                                                                            &                                                                                                                       \\ \cline{1-5}
8                       & \multicolumn{1}{c|}{\begin{tabular}[c]{@{}c@{}} $1.0 \pm 0.0$\end{tabular}}                         & \multicolumn{1}{c|}{\begin{tabular}[c]{@{}c@{}} $0.99 \pm 0.095$\end{tabular}} & \multicolumn{1}{c|}{\begin{tabular}[c]{@{}c@{}} $0.979 \pm 0.092$\end{tabular}}  & \multicolumn{1}{c|}{\begin{tabular}[c]{@{}c@{}} $0.944 \pm 0.18$\end{tabular}}   &                                                                                                                                            &                                                                                                                                            &                                                                                                                                            &                                                                                                                                          &                                                                                                                                            &                                                                                                                       \\ \cline{1-6}
10                      & \multicolumn{1}{c|}{\begin{tabular}[c]{@{}c@{}} $1.0 \pm 0.0$\end{tabular}}                                & \multicolumn{1}{c|}{\begin{tabular}[c]{@{}c@{}} $1.0 \pm 0.0$\end{tabular}}    & \multicolumn{1}{c|}{\begin{tabular}[c]{@{}c@{}} $0.997 \pm 0.021$\end{tabular}} & \multicolumn{1}{c|}{\begin{tabular}[c]{@{}c@{}} $0.997 \pm 0.023$\end{tabular}}   & \multicolumn{1}{c|}{\begin{tabular}[c]{@{}c@{}} $0.985 \pm 0.052$\end{tabular}} &                                                                                                                                            &                                                                                                                                            &                                                                                                                                          &                                                                                                                                            &                                                                                                                       \\ \cline{1-7}
12                      & \multicolumn{1}{c|}{\begin{tabular}[c]{@{}c@{}} $1.0 \pm 0.0$\end{tabular}}                            & \multicolumn{1}{c|}{\begin{tabular}[c]{@{}c@{}} $1.0 \pm 0.0$\end{tabular}}    & \multicolumn{1}{c|}{\begin{tabular}[c]{@{}c@{}} $1.0 \pm 0.0$\end{tabular}}    & \multicolumn{1}{c|}{\begin{tabular}[c]{@{}c@{}} $0.996 \pm 0.036$\end{tabular}} & \multicolumn{1}{c|}{\begin{tabular}[c]{@{}c@{}} $0.989 \pm 0.097$\end{tabular}}  & \multicolumn{1}{c|}{\begin{tabular}[c]{@{}c@{}} $0.985 \pm 0.055$\end{tabular}} &                                                                                                                                            &                                                                                                                                          &                                                                                                                                            &                                                                                                                       \\ \cline{1-8}
14                      & \multicolumn{1}{c|}{\begin{tabular}[c]{@{}c@{}} $1.0 \pm 0.0$\end{tabular}}                                 & \multicolumn{1}{c|}{\begin{tabular}[c]{@{}c@{}} $1.0 \pm 0.0$\end{tabular}}        & \multicolumn{1}{c|}{\begin{tabular}[c]{@{}c@{}} $0.997 \pm 0.031$\end{tabular}} & \multicolumn{1}{c|}{\begin{tabular}[c]{@{}c@{}} $1.0 \pm 0.0$\end{tabular}}      & \multicolumn{1}{c|}{\begin{tabular}[c]{@{}c@{}} $1.0 \pm 0.0$\end{tabular}}     & \multicolumn{1}{c|}{\begin{tabular}[c]{@{}c@{}} $0.999 \pm 0.006$\end{tabular}}  & \multicolumn{1}{c|}{\begin{tabular}[c]{@{}c@{}} $0.986 \pm 0.063$\end{tabular}} &                                                                                                                                          &                                                                                                                                            &                                                                                                                       \\ \cline{1-9}
16                      & \multicolumn{1}{c|}{\begin{tabular}[c]{@{}c@{}} $1.0 \pm 0.0$\end{tabular}}                                & \multicolumn{1}{c|}{\begin{tabular}[c]{@{}c@{}} $1.0 \pm 0.0$\end{tabular}}         & \multicolumn{1}{c|}{\begin{tabular}[c]{@{}c@{}} $1.0 \pm 0.0$\end{tabular}}     & \multicolumn{1}{c|}{\begin{tabular}[c]{@{}c@{}} $1.0 \pm 0.0$\end{tabular}}     & \multicolumn{1}{c|}{\begin{tabular}[c]{@{}c@{}} $1.0 \pm 0.0$\end{tabular}}     & \multicolumn{1}{c|}{\begin{tabular}[c]{@{}c@{}} $0.997 \pm 0.027$\end{tabular}}   & \multicolumn{1}{c|}{\begin{tabular}[c]{@{}c@{}} $0.999 \pm 0.006$\end{tabular}}  & \multicolumn{1}{c|}{\begin{tabular}[c]{@{}c@{}} $0.993 \pm 0.02$\end{tabular}}  &                                                                                                                                            &                                                                                                                       \\ \cline{1-10}
18                      & \multicolumn{1}{c|}{\begin{tabular}[c]{@{}c@{}} $1.0 \pm 0.0$\end{tabular}}                                & \multicolumn{1}{c|}{\begin{tabular}[c]{@{}c@{}} $1.0 \pm 0.0$\end{tabular}}         & \multicolumn{1}{c|}{\begin{tabular}[c]{@{}c@{}} $1.0 \pm 0.0$\end{tabular}}    & \multicolumn{1}{c|}{\begin{tabular}[c]{@{}c@{}} $1.0 \pm 0.0$\end{tabular}}      & \multicolumn{1}{c|}{\begin{tabular}[c]{@{}c@{}} $0.999 \pm 0.012$\end{tabular}}  & \multicolumn{1}{c|}{\begin{tabular}[c]{@{}c@{}} $1.0 \pm 0.005$\end{tabular}}    & \multicolumn{1}{c|}{\begin{tabular}[c]{@{}c@{}} $1.0 \pm 0.0$\end{tabular}}     & \multicolumn{1}{c|}{\begin{tabular}[c]{@{}c@{}} $1.0 \pm 0.001$\end{tabular}} & \multicolumn{1}{c|}{\begin{tabular}[c]{@{}c@{}} $0.995 \pm 0.016$\end{tabular}} &                                                                                                                       \\ \hline
20                      & \multicolumn{1}{c|}{\begin{tabular}[c]{@{}c@{}} $1.0 \pm 0.0$\end{tabular}}                                & \multicolumn{1}{c|}{\begin{tabular}[c]{@{}c@{}} $1.0 \pm 0.0$\end{tabular}}        & \multicolumn{1}{c|}{\begin{tabular}[c]{@{}c@{}} $1.0 \pm 0.0$\end{tabular}}     & \multicolumn{1}{c|}{\begin{tabular}[c]{@{}c@{}} $1.0 \pm 0.0$\end{tabular}}     & \multicolumn{1}{c|}{\begin{tabular}[c]{@{}c@{}} $1.0 \pm 0.0$\end{tabular}}       & \multicolumn{1}{c|}{\begin{tabular}[c]{@{}c@{}} $1.0 \pm 0.0$\end{tabular}}     & \multicolumn{1}{c|}{\begin{tabular}[c]{@{}c@{}} $1.0 \pm 0.003$\end{tabular}}    & \multicolumn{1}{c|}{\begin{tabular}[c]{@{}c@{}} $1.0 \pm 0.0$\end{tabular}}    & \multicolumn{1}{c|}{\begin{tabular}[c]{@{}c@{}} $0.999 \pm 0.005$\end{tabular}} & \begin{tabular}[c]{@{}c@{}} $0.992 \pm 0.025$\end{tabular} \\ \hline
\end{tabular}
\label{tab:phase_alignment_random}
\end{table*}

To assess the performance of the \texttt{GreedySAT}, we benchmark it against the \texttt{OPT} strategy in producing the highest $g_{\alpha}$ [Eq. (\ref{eq:truegrad_app})] for a fixed $\hat X_{\alpha}$, over a class of uniformly random phase-alignment problems. Such problems are obtained by the following prescription:
\begin{itemize}
\item{The $L$ configurations $\{ \ket{\phi_i} \}_{i=1}^{L}$ defining the model space are independently uniformly sampled from the set of $N_q$-bit strings, $\{ 0, 1 \}^{\otimes N_q}$.}
\item{Similarly, the $\hat X_{\alpha}$'s binary representation, $\vec{\mu}^{(\alpha)}$, is also uniformly sampled from the set of $N_q$-bit strings.}
\item{The values of $\{ \Xi^{(\alpha)}_{i} \}_{i=1}^{L}$ are independently uniformly sampled from within the interval $[-1, 1]$.}
\end{itemize}
To denote a specific instance, we label the problem by $g^{\rm{SAT}/\rm{OPT}}_{\alpha}(\{ \ket{\phi_i} \}, \{ \Xi^{(\alpha)}_{i} \})$ where the superscript refers to which of $\rm{GreedySAT}$ and $\rm{OPT}$ was used. To quantify the performance we take their ratios,
\begin{align*}
R_\alpha(\{ \ket{\phi_i} \}, \{ \Xi^{(\alpha)}_{i} \}) & = \frac{g_{\alpha}^{\rm{SAT}}(\{ \ket{\phi_i} \}, \{ \Xi^{(\alpha)}_{i} \})}{g_{\alpha}^{\rm{OPT}}(\{ \ket{\phi_i} \}, \{ \Xi^{(\alpha)}_{i} \})}.
\end{align*}
By performing many uniformly random samples of the phase-alignment problem, we empirically approximate the average case ratios:
\begin{align*}
R_{avg} = \mathbb{E}_{\{ \ket{\phi_i} \}, \{ \Xi^{(\alpha)}_{i} \}, \vec{\mu}^{(\alpha)}}\  R_\alpha(\{ \ket{\phi_i} \}, \{ \Xi^{(\alpha)}_{i} \}).
\end{align*}
In Table \ref{tab:phase_alignment_random} we report the values of $R_{avg} $ for a range of phase-alignment problem instances. Each reported mean is obtained from $100$ uniformly sampled phase-alignment problems of $L$-dimensional model spaces defined on $N_q$ qubits. Since MS-iQCC does not require large model spaces for the multiconfigurational reference states, we restrict $L \leq N_q$ in this analysis. The mean of means of ratios obtained across all $(N_q, L)$'s considered is $0.995 \pm 0.009$, i.e., it is nearly indistinguishable in average performance from the \texttt{OPT} solution. While the performance may begin to suffer in the $L > N_q$ regime, where the number of equations is larger than number of free variables, this is generally not considered problematic in the context of MS-iQCC where relatively simple CSFs may be used as reference states, leading to small model spaces.

\end{document}